\documentclass[aps,reprint]{revtex4-2}

\usepackage{amsmath,bm}
\usepackage{amssymb}
\usepackage{mathrsfs}
\usepackage{amsfonts}
\usepackage{graphicx}
\usepackage{siunitx}
\usepackage[version=3]{mhchem}
\usepackage[acronym]{glossaries}
\usepackage{booktabs}

\DeclareSIUnit\au{\text {au}}

\newacronym{cboa}{CBOA}{cavity Born-Oppenheimer approximation}
\newacronym{bo}{BOA}{Born-Oppenheimer approximation}
\newacronym{cbohf}{CBO-HF}{cavity Born-Oppenheimer Hartree-Fock}
\newacronym{scf}{SCF}{self-consistent field}
\newacronym{dse}{DSE}{dipole self-energy}
\newacronym{pes}{PES}{potential energy surface}
\newacronym{cpes}{cPES}{cavity potential energy surface}
\newacronym{vsc}{VSC}{vibrational-strong coupling}
\newacronym{esc}{ESC}{electronic-strong coupling}
\newacronym{lp}{LP}{lower polariton}
\newacronym{up}{UP}{upper polariton}
\newacronym{ejcm}{EJCM}{extended molecular Jaynes-Cummings model}

\newacronym{sd}{SD}{Steepest Descent}
\newacronym{nr}{NR}{Newton–Raphson}
\newacronym{bf}{BFGS}{Broyden–Fletcher–Goldfarb–Shanno}
\newacronym{bfi}{BFGS-I}{Broyden-Fletcher-Goldfarb-Shanno initiated with the exact Hessian}
\newacronym{bfhx}{BFGS-Hx}{Broyden-Fletcher-Goldfarb-Shanno with recalculated Hessian every xth step}
\newacronym{bfh10}{BFGS-H10}{Broyden-Fletcher-Goldfarb-Shanno with recalculated Hessian every tenth step}

\begin{document}

\title{Do Molecular Geometries Change Under Vibrational Strong Coupling?}

\author{Thomas Schnappinger}
\email{thomas.schnappinger@fysik.su.se}
\affiliation{Department of Physics, Stockholm University, AlbaNova University Center, SE-106 91 Stockholm, Sweden}

\author{Markus Kowalewski}
\email{markus.kowalewski@fysik.su.se}
\affiliation{Department of Physics, Stockholm University, AlbaNova University Center, SE-106 91 Stockholm, Sweden}
\date{\today}%

\begin{abstract}
As pioneering experiments have shown, strong vibrational coupling between molecular vibrations and light modes in an optical cavity can significantly alter molecular properties and even affect chemical reactivity. However, the current theoretical description is limited and far from complete. To explore the origin of this exciting observation, we investigate how the molecular structure changes under strong light-matter coupling using an \textit{ab-initio} method based on the cavity Born-Oppenheimer Hartree-Fock ansatz. By optimizing \ce{H2O} and \ce{H2O2} resonantly coupled to cavity modes, we study the importance of reorientation and geometric relaxation. In addition, we show that the inclusion of one or two cavity modes can change the observed results. On the basis of our findings, we derive a simple concept to estimate the effect of the cavity interaction on the molecular geometry using the molecular polarizability and the dipole moments.
\end{abstract}

\maketitle
\section{Introduction}

When molecules are placed in a non-classical photonic environment present in optical or nanoplasmonic cavities, it is possible to form strong light-matter-coupled hybrid states called polaritons~\cite{Benz2016-vq,Griffiths2021-ge,Basov2021-hr,Ribeiro2018-xd,Ebbesen2023-fd,Bhuyan2023-se}.
The control of the photonic environment allows to couple the cavity photon modes to vibrational or electronic transitions in molecules, called \gls{vsc} or \gls{esc}, respectively.
Both types of strong coupling can be an effective tool for modifying molecular properties and offer a possible novel approach to control chemical reactions using optical resonators.
The experimental advances reported cover a wide range of applications, from manipulating the selectivity of organic reactions~\cite{Hutchison2012-od,Thomas2019-ve,Sau2021-vb}, changing the ionic conductivity of water~\cite{Fukushima2022-ox} to even influencing enzymatic activity~\cite{Vergauwe2019-wt,Bai2023-ev}.
Driven by these experimental advances, considerable efforts have been made to develop theories that elucidate the mechanisms governing polaritonic chemistry.
Even so, the current theoretical description is limited and far from complete.
However, in recent years, a substantial number of studies using different theoretical approaches have proposed that a variety of additional reactions can be enhanced, inhibited, or controlled~\cite{Felicetti2020-qq,Campos-Gonzalez-Angulo2020-ql,Li2021-ri,Galego2019-xg,Schafer2022-id,Li2020-bz,Pavosevic2023-ik,Pavosevic2023-vc,Vidal2022-sy,Riso2022-du,Severi2023-qz,Vu2022-mx,Pavosevic2022-oy}.
In particular, the combination of electronic structure methods and quantum electrodynamics~\cite{Tokatly2013-ho,Ruggenthaler2014-it,Flick2017-jh,Haugland2020-xh,McTague2022-xl,Schnappinger2023-hh,Foley2023-pq,Angelico2023-ll,Liebenthal2023-ra} has significantly improved theoretical understanding and will hopefully help close the existing gaps between theory and experiment.
Most of these studies model cavity-induced electronic structure changes using a single molecule coupled to a single-cavity mode in the strong-coupling limit.
In almost every example, both a fixed orientation relative to the polarization axes of the cavity mode and fixed molecular geometries are assumed, despite the fact that the coupling strengths used are quite large.

In this work, we use  the \gls{cbohf} ansatz~\cite{Schnappinger2023-hh} together with analytical gradients~\cite{Schnappinger2023-wp} to optimize \ce{H2O} and \ce{H2O2} resonantly coupled to one and two cavity photon modes.
The \gls{cbohf} ansatz is capable of describing the electronic ground state of single molecules, as well as an ensemble of molecules coupled to an optical cavity under \glspl{vsc} conditions~\cite{Schnappinger2023-hh,Angelico2023-ll,Sidler2024-vm}.
By performing the optimizing the geometries in the laboratory frame together with the polarization vectors of the cavity, we are able to study both the orientation and the relaxation of the internal coordinates of the molecules induced by the interaction with the cavity photon modes.
Furthermore, we calculate vibro-polaritonic IR spectra~\cite{Schnappinger2023-wp} within the harmonic approximation, we allow for the  verification of the structures found as real minima and to
analyze in detail the formed polaritonic states.
As will become clear in the remainder of this article, without a physical mechanism to fix the orientation, molecules inside an optical cavity will orient and change their geometry depending on the coupling strength.
Furthermore, these observed effects will vary if one- or two-cavity orthogonal photon modes are included in the simulation.
The main results of this work are indication that some theoretical studies in the literature may overestimate the cavity-induced effects on the ground-state chemistry.
Finally, we establish a useful and straightforward connection between the molecular polarizability and dipole moment and the expected reorientation and relaxation of a molecule coupled to an optical cavity.

\section{Results \& Discussion}
\subsection{Optimization of \ce{H2O} embedded in a cavity}

\begin{figure}
     \centering
         \includegraphics[width=0.49\textwidth]{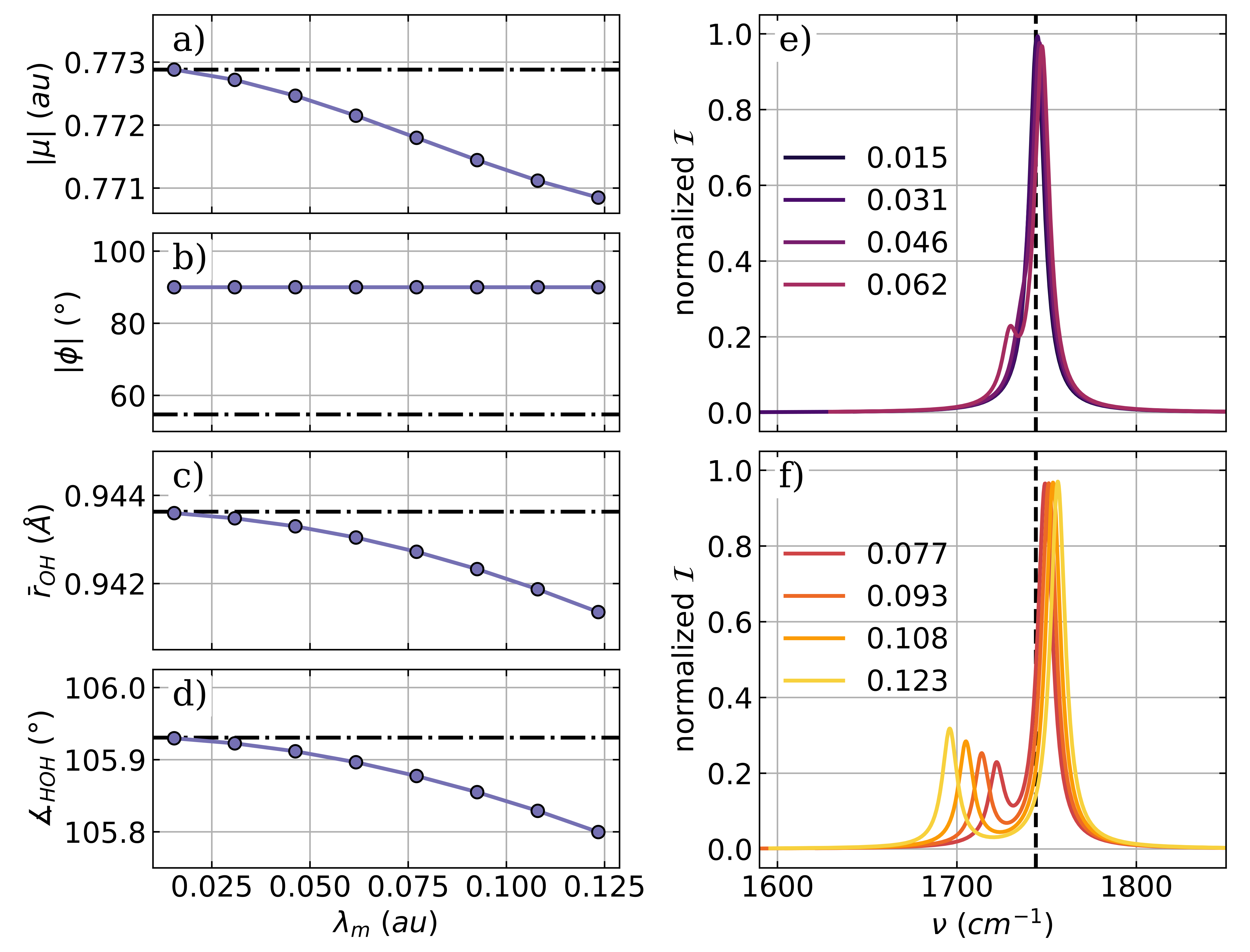}
    \caption{Optimized parameters of a \ce{H2O} molecule coupled to a single photon mode of an optical cavity as a function of the coupling strength $\lambda_m$. a) Magnitude of the dipole moment $|\mu|$ b) the angle $\phi$ between the polarization axis of the cavity and the dipole moment, c) averaged \ce{OH} bond length and d) bond angle. The dashed-dotted lines in a)-d) indicate the initial values. Relevant parts of the vibro-polaritonic IR spectra for different coupling strengths (color-coded) are shown in e) and f). The cavity frequency $\omega_m$ is \SI{1744}{\per\centi\meter} shown as a black dashed line in e) and f). The cavity coupling $\lambda_{m}$ increases from \SI{0.015}{\au} to \SI{0.123}{\au} and the cavity polarization axis is $\bm{e} = \left(1,0,0\right)$.}
\label{fig:opt_h2o_1}
\end{figure}

As a first example, we optimize a \ce{H2O} molecule coupled to one photon mode of an optical cavity.
The cavity frequency $\omega_m$ is resonant with the bending mode, which has a field-free vibrational frequency of \SI{1744}{\per\centi\meter}, and the cavity polarization axis $\bm{e}$ is aligned with the $x$ axis of the laboratory frame.
Fig.~\ref{fig:opt_h2o_1} shows the optimized parameters for the coupled molecular cavity system as a function of the coupling strength $\lambda_m$ and the corresponding vibro-polaritonic IR spectra in the region of the \ce{H2O} bending mode.

For the optimized single-mode cavity-\ce{H2O} system, the magnitude of the dipole $|\langle \mu \rangle|$ moment, shown in Fig.~\ref{fig:opt_h2o_1}~a), is only slightly reduced with increasing $\lambda_m$.
This rather small change is consistent with the observed small decrease in both the \ce{OH} bond length (Fig.~\ref{fig:opt_h2o_1}~c)) and bond angle (Fig.~\ref{fig:opt_h2o_1}~d)) even at high coupling strengths.
The only system parameter significantly affected by \gls{vsc} is the relative orientation of the molecular dipole moment and the polarization axis of the cavity mode visualized as the angle $\phi$ between them in Fig.~\ref{fig:opt_h2o_1}~b).
For the arbitrary chosen initial configuration, $\phi$ has a value of about \ang{55} (dashed dotted line), which changes to exactly \ang{90} after optimization, regardless of the coupling strength.
The interaction with teh photon field leads to an orientation of the molecule so that not only the dipole moment but also the molecular plan is orthogonal to the polarization axis.
The relevant parts of the vibro-polaritonic IR spectra for the optimized water-cavity system are shown in Fig.~\ref{fig:opt_h2o_1}~e) and~f) for different coupling strengths.
Given the orientation of the dipole moment perpendicular to the cavity polarization axis, one would expect no hybrid photonic states to be formed because of the lack of dipole-cavity interaction.
This assumption holds for lower coupling strengths below \SI{0.062}{\au}, shown in Fig.~\ref{fig:opt_h2o_1}~e).
Within the used broadening of \SI{10}{\per\centi\meter}, only a single peak is observed, almost unshifted to the field-free value (black dashed line).
For $\lambda_m= \SI{0.062}{\au}$ a small shoulder appears at a slightly lower frequency, the bright purple line in Fig.~\ref{fig:opt_h2o_1}~e).
If the coupling is further increased, as shown in Fig.~\ref{fig:opt_h2o_1}~f), this shoulder becomes a separate peak with increasing intensities and shifts to lower frequencies with increasing $\lambda_m$.
As introduced in our early work~\cite{Schnappinger2023-wp}, the normal mode component that describes the change in the classical photon displacement field is a measure of how much photon character the corresponding transition is.
The main peak in all shown vibro-polaritonic IR spectra has no photonic character and can be described as a pure molecular transition corresponding to the bending mode.
The smaller peak at higher coupling strengths is predominantly photonic, and we do not see the formation of a typical hybrid matter-photon \gls{lp} state and \gls{up} state here.
This can be explained by a rotational motion of the \ce{H2O} molecule.
Due to the confinement of the cavity, this motion is not a free rotation anymore.
This motion creates a dipole moment parallel to the cavity polarization axis,
which leads to a coupling with the photon mode.
Due to this coupling, the photonic transition gains intensity and shifts to lower frequencies.
More details and a thorough analysis of the photonic characters of all relevant transitions can be found in the Supporting Information.

\begin{figure}
     \centering
         \includegraphics[width=0.49\textwidth]{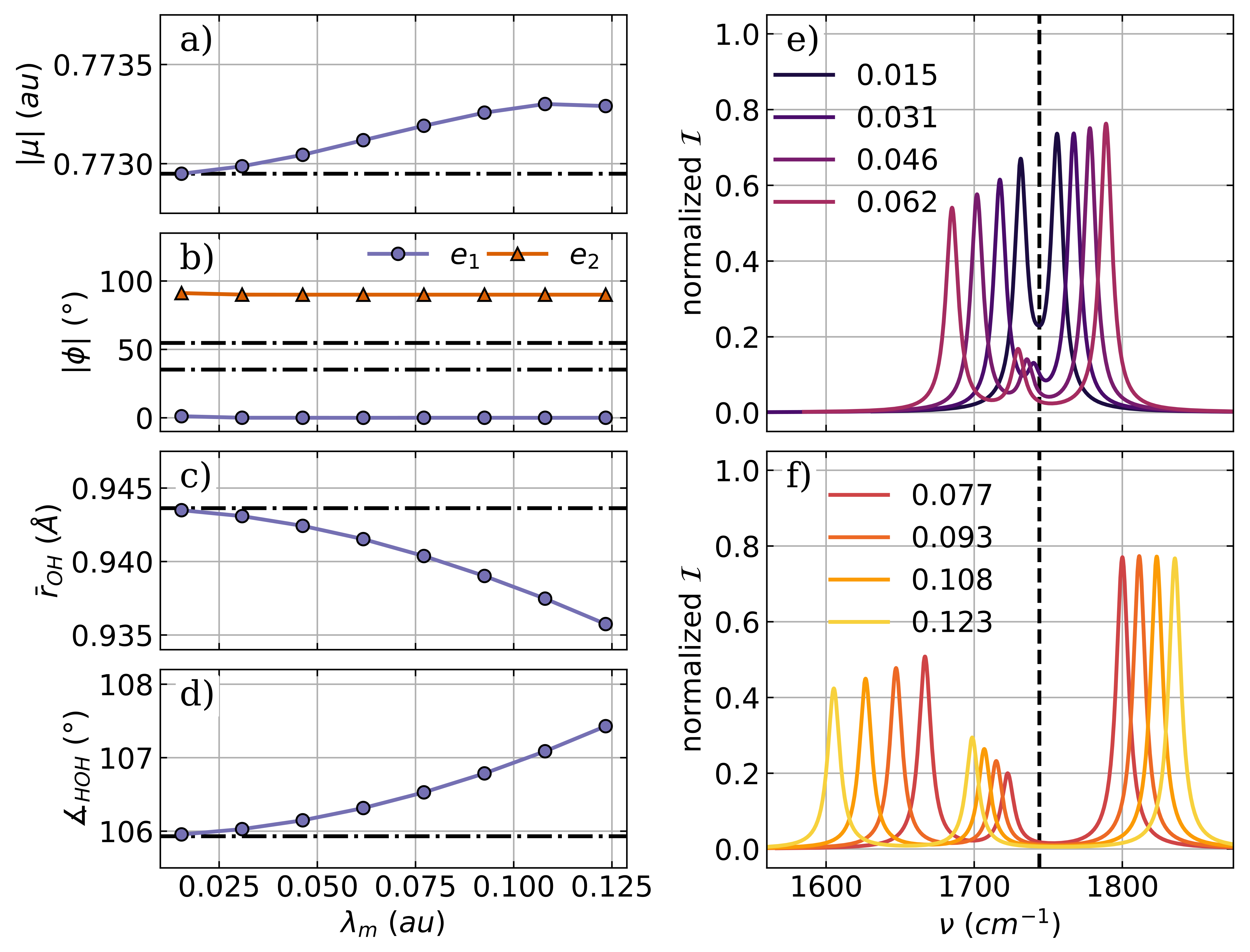}
    \caption{Optimized parameters of a \ce{H2O} molecule coupled to two orthogonal cavity photon modes as a function of the coupling strength $\lambda_m$. a) Magnitude of the dipole moment $|\mu|$ b) the two angle $\phi$ between the polarization axes of the cavity and the dipole moment of the molecule, c) averaged \ce{OH} bond length and d) the bond angle. The dashed-dotted lines indicate the initial values in a)-d). The relevant part of the vibro-polaritonic IR spectra for different coupling strengths (color-coded) is shown in e) and f). The cavity frequency $\omega_m$ is \SI{1744}{\per\centi\meter} shown as a black dashed line in e) and f). The cavity coupling $\lambda_{m}$ increases from \SI{0.015}{\au} to \SI{0.123}{\au} and the cavity polarization axes are $\bm{e}_1 = \left(1,0,0\right)$ and $\bm{e}_2 = \left(0,1,0\right)$.}
\label{fig:opt_h2o_2}
\end{figure}

Next, we discuss the optimization results for \ce{H2O} coupled to two cavity modes of orthogonal polarization with the same frequency, effectively modeling a Fabry-P\'erot-like setup.
In practice, we use the same cavity frequency and coupling strength for both modes and take the same values as for the single-mode case.
The polarization axes are aligned with the $x$ axis ($\bm{e}_1$) and the $y$ axis ($\bm{e}_2$) of the laboratory
frame.
The optimized parameters as a function of the coupling strength $\lambda_m$ and the corresponding vibro-polaritonic IR spectra in the region of the \ce{H2O} bending mode are shown in Fig.~\ref{fig:opt_h2o_2}

When the optimized parameters are compared for the single-mode case and the two-mode case, some similarities but also distinct differences are observed.
The change in magnitude of the dipole moment, shown in Fig.~\ref{fig:opt_h2o_2}~a), is nearly one order of magnitude smaller for the two-mode optimization.
In contrast, the \ce{OH} bond length (Fig.~\ref{fig:opt_h2o_2}~c)) and the bond angle (Fig.~\ref{fig:opt_h2o_2}~d)) are more strongly influenced by the coupling to two cavity modes.
In particular, the change in the bond angle is significantly larger compared to the case of a single mode, although it increases only by \ang{1.5} for the highest coupling strength.
Again, the most significant changes due to optimization are the relative orientations of the molecular dipole moment with respect to the cavity-mode polarization axes.
The corresponding angles $\phi$ for $\bm{e}_1$ and $\bm{e}_2$ are shown in Fig.~\ref{fig:opt_h2o_2}~b).
Starting from a configuration where both $\bm{e}_1$ and $\bm{e}_2$ are in the molecular plane but not aligned with the dipole moment, the optimization reorients the molecule independently of the coupling strength so that $\bm{e}_1$ is parallel to the dipole moment, while $\bm{e}_2$ is orthogonal to the dipole moment and the molecular plane.
The corresponding vibro-polaritonic IR spectra for the optimized water-cavity system are shown in Fig.~\ref{fig:opt_h2o_2}~e) and~f) for different coupling strengths.
Already, for the lowest coupling strength $\lambda_m=0.015 au$ in Fig.~\ref{fig:opt_h2o_2}~e), a splitting into a \gls{lp} transition and a \gls{up} transition is observed.
The corresponding normal modes clearly show a hybridization of the vibrational transition and the cavity photon mode with the polarization axis $\bm{e}_1$.
As $\lambda_m$ increases, the Rabi splitting between the \gls{lp} and \gls{up} transitions increases and becomes more asymmetric, consistent with our previous work~\cite{Schnappinger2023-wp,Sidler2024-vm}.
At the same time, a third weaker peak between the \gls{lp} and \gls{up} transitions becomes visible.
Similarly to the single-mode case, this peak is due to the coupling between a specific constricted rotational degree of freedom and the photon mode with the polarization axis $\bm{e}_2$.
Due to this coupling, the photonic transition gains intensity and shifts to low frequencies with increasing coupling strength.
More information on the photonic characters of all relevant transitions can be found in the Supporting Information.

\begin{figure}
     \centering
         \includegraphics[width=0.48\textwidth]{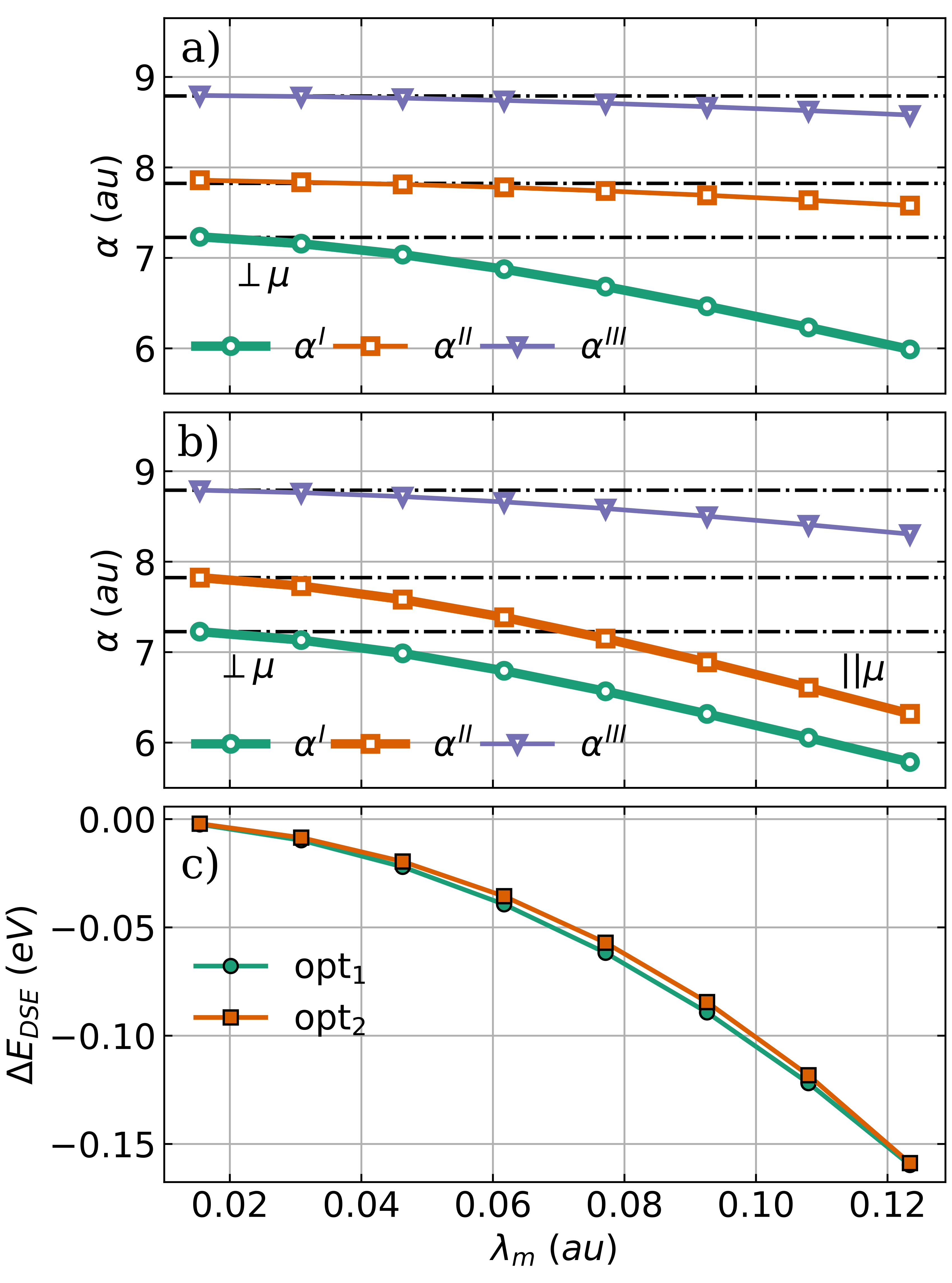}
    \caption{Principal components of the polarizability tensor $\alpha$ as a function of the coupling strength $\lambda_m$ for optimized \ce{H2O} structures coupled to a) a single cavity mode and to b) two orthogonal cavity modes. The bold line indicates the component aligned with the cavity polarization axes, and the black dashed dotted lines represent the field-free principal components of the polarizability. c) Change in \gls{dse} contribution as a function of the coupling strength $\lambda_m$ compared to the initial \ce{H2O} geometry for the single-mode case (green) and the two-mode case (orange). The cavity frequency $\omega_m$ is \SI{1744}{\per\centi\meter}.}
\label{fig:alpha_h2o}
\end{figure}

To gain more insight into the underlying driving force for the energetically favorable
orientation, we discuss how \gls{vsc} modifies the polarizability $\alpha$ and the \gls{dse} contribution.
The principal components of the polarizability tensor $\alpha$ and the change in the \gls{dse} contribution are shown in Fig.~\ref{fig:alpha_h2o} as a function of the coupling strength for both optimized \ce{H2O}-cavity systems.

In the single-mode case shown in Fig.~\ref{fig:alpha_h2o}~a) and the two-mode case shown in Fig.~\ref{fig:alpha_h2o}~b) all three principal components of $\alpha$ are reduced with increasing coupling strength or, in other words, the interaction with the photon mode contracts the electronic density.
Therefore, it is logical that the components aligned with the cavity polarization axes (highlighted in bold) are the most affected.
These components already have the smallest value in the field-free case, and the optimization reorients the molecule so that these components align with the cavity polarization axes.
For \ce{H2O} these are the components orthogonal to the molecular plane corresponding to the green lines in Fig.~\ref{fig:alpha_h2o}~a) and b) and the one parallel to the molecular dipole moment corresponding to the orange lines in Fig.~\ref{fig:alpha_h2o}~a) and b).
As shown in Fig.~\ref{fig:alpha_h2o}~c), the observed reorientation also minimizes the \gls{dse} contribution to the coupled cavity-molecule system.
This is in agreement with our previous work~\cite{Schnappinger2023-hh} and a recent study by Liebenthal and DePrince~\cite{Liebenthal2024-xb}.
In general, reorientation can be rationalized as an effective way to reduce molecular polarizability along the cavity polarization axes and thus minimizing the \gls{dse} contribution.

All results presented for the optimization of \ce{H2O} resonantly coupled to a single or two orthogonal cavity photon modes show that, without restriction of the rotational degrees of freedom, the cavity interaction leads to a reorientation of the molecule.
This rotational motion occurs for all coupling strengths and is more pronounced than changes in the internal coordinates of \ce{H2O}.
For higher coupling strengths, the rotational effects even become visible in the vibro-polaritonic IR spectra.

\subsection{Optimization of \ce{H2O2} embedded in a cavity}

\begin{figure}
     \centering
         \includegraphics[width=0.49\textwidth]{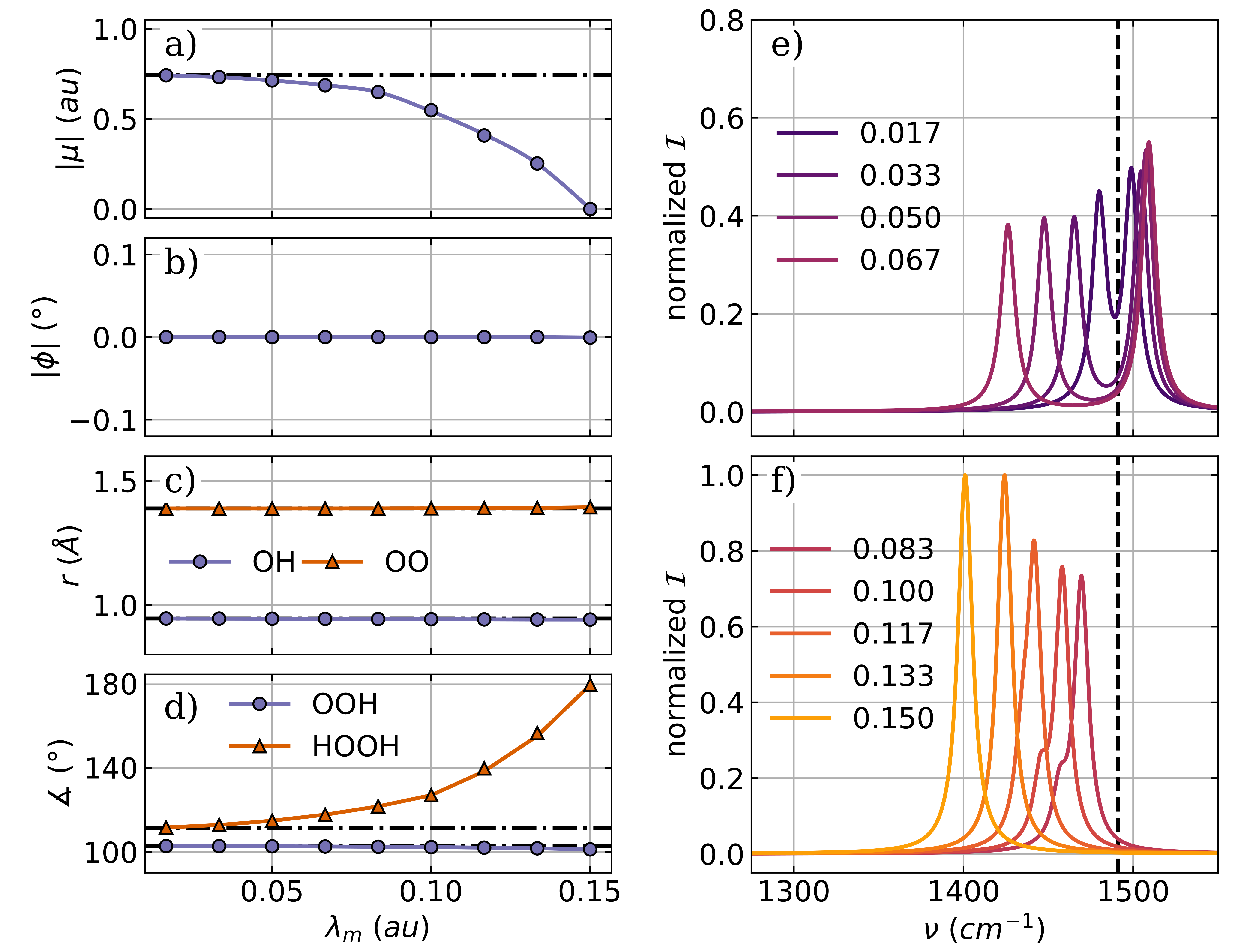}
    \caption{Optimized parameters of a \ce{H2O2} molecule coupled to a single photon mode of an optical cavity as a function of the coupling strength $\lambda_m$. a) Magnitude of the dipole moment $|\mu|$ b) the angle $\phi$ between the polarization axis of the cavity and the dipole moment of the molecule, c) averaged \ce{OH} bond length (purple) and \ce{OO} bond length (orange) and d) the averaged bond angle (purple) as well as the dihedral angle (orange). The dashed-dotted lines in a)-d) indicate the initial values. Relevant parts of the vibro-polaritonic IR spectra for different coupling strengths (color-coded) are shown in e) and f). The cavity frequency $\omega_m$ is \SI{1491}{\per\centi\meter} shown as a black dashed line in e) and f). The cavity coupling $\lambda_{m}$ increases from \SI{0.017}{\au} to \SI{0.150}{\au} and the cavity polarization axis is $\bm{e} = \left(0,0,1\right)$.}
\label{fig:opt_h2o2_1}
\end{figure}

The optimization of a \ce{H2O2} molecule coupled to a single photon mode and two to orthogonal photon modes serves as the second example.
The cavity is resonant with the asymmetric bending mode of \SI{1491}{\per\centi\meter}, and the cavity polarization axis $\bm{e}$ in the case of a single mode is aligned with the molecular dipole moment, which corresponds to the $z$ axis of the laboratory frame.
The optimized parameters for the coupled molecular cavity system as a function of the coupling strength $\lambda_m$ and the relevant part of the corresponding vibro-polaritonic IR spectra are shown in Fig.~\ref{fig:opt_h2o2_1}.

Regardless of the coupling strength used, the dipole moment of \ce{H2O2} remains aligned with the polarization axis of the single cavity mode, as shown in Fig.~\ref{fig:opt_h2o2_1}~b).
However, its size decreases significantly with increasing coupling up to the situation where the dipole moment is zero at $\lambda_m = \SI{0.150}{\au}$ (see Fig.~\ref{fig:opt_h2o2_1}~a)).
This large change in the dipole moment with the increasing coupling strength is induced by an increase in the dihedral angle, leading to a planarization of \ce{H2O2} (see orange line Fig.~\ref{fig:opt_h2o2_1}~d)).
For $\lambda_m = \SI{0.150}{\au}$ the \ce{H2O2} molecule becomes completely planar in a \textit{trans} configuration, which results in a zero dipole moment.
Note that for this \textit{trans} structure, the cavity polarization axis is orthogonal with respect to the molecular plane, very similar to the \ce{H2O} case.
In contrast, the bond lengths and the averaged bond angle shown in Fig.~\ref{fig:opt_h2o2_1}~c) and Fig.~\ref{fig:opt_h2o2_1}~d), respectively, remain nearly constant for increasing couplings strengths.
The observed change in the molecular geometry due to the cavity interaction can be clearly seen in the corresponding vibro-polaritonic IR spectra, shown in Fig.~\ref{fig:opt_h2o2_1}~e) and~f) for different coupling strengths.
Details about the photonic characters of all relevant transitions can be found in the Supporting Information.
For smaller values of $\lambda_m$, as depicted in Fig.~\ref{fig:opt_h2o2_1}~e), there is a clear splitting into a \gls{lp} transition and a \gls{up} transition, which increases with increasing coupling strength.
If $\lambda_m$ is further increased, the changes in dipole moment and dihedral angle are more pronounced, which changes the vibro-polaritonic IR spectra, see Fig.~\ref{fig:opt_h2o2_1}~f).
These spectra are characterized by a single peak that is a pure molecular transition but is red-shifted relative to the field-free asymmetric bending mode.
For a coupling strength below \SI{0.13}{\au} a small shoulder a slightly lower frequency is present.
This signal has a photonic character and is visible due to the weak coupling of the photon mode with the twisting mode in \ce{H2O2} at \SI{424}{\per\centi\meter}.
This mode is the most intense transition in the field-free spectrum and is characterized by a change in the dihedral angle of \ce{H2O2}.
Similarly to the rotational motion observed in \ce{H2O}, this twisting mode induces a dipole moment parallel to the polarization axis of the cavity.
However, as the coupling increases further and \ce{H2O2} becomes more planar, this coupling weakens, and only the single molecular peak remains visible.

\begin{figure}
     \centering
         \includegraphics[width=0.48\textwidth]{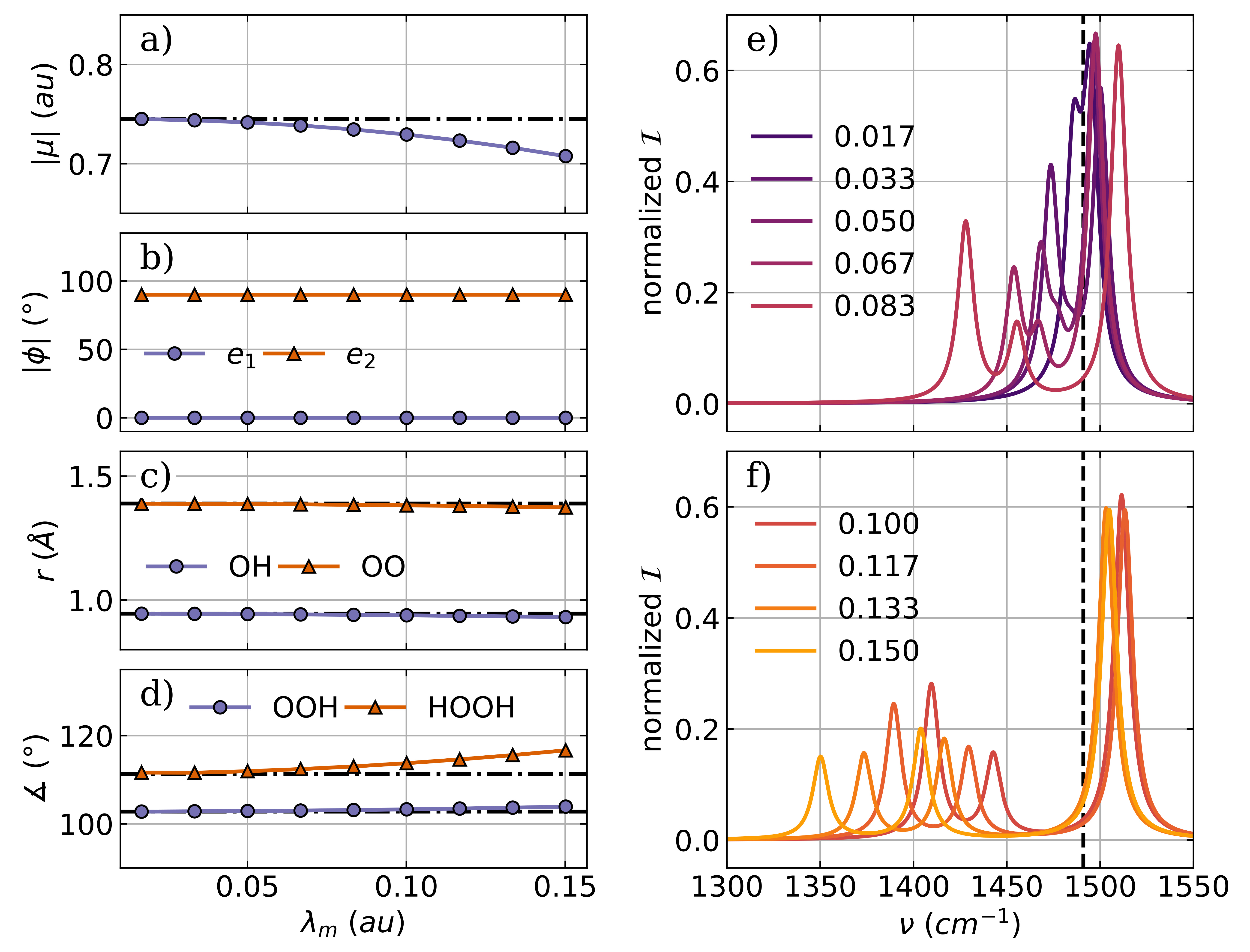}
    \caption{Optimized parameters of a \ce{H2O2} molecule coupled to two orthogonal cavity photon modes as a function of the coupling strength $\lambda_m$. a) Magnitude of the dipole moment $|\mu|$ b) the angle $\phi$ between the cavity polarization axes and the dipole moment of the molecule, c) averaged \ce{OH} bond length (purple) and \ce{OO} bond length (orange) and d) the averaged bond angle (purple) as well as the dihedral angle (orange). The dashed-dotted lines in a)-d) indicate the initial values. Relevant parts of the vibro-polaritonic IR spectra for different coupling strengths (color-coded) are shown in e) and f). The cavity frequency $\omega_m$ is \SI{1491}{\per\centi\meter} shown as a black dashed line in e) and f). The cavity coupling $\lambda_{m}$ increases from \SI{0.017}{\au} to \SI{0.150}{\au} and the cavity polarization axes are $\bm{e}_1 = \left(0,0,1\right)$ and $\bm{e}_2 = \left(1,0,0\right)$.}
\label{fig:opt_h2o2_2}
\end{figure}

To optimize the \ce{H2O2} molecule coupled to a two mode cavity, we add a second orthogonal cavity mode with the same frequency and coupling strength.
The two polarization axes are aligned with the $z$ axis ($\bm{e}_1$) and the $x$ axis ($\bm{e}_2$) of the laboratory frame.
The optimized parameters as a function of the coupling strength $\lambda_m$ and the relevant part of the vibro-polaritonic IR spectra are shown in Fig.~\ref{fig:opt_h2o2_2}

Optimization of \ce{H2O2} coupled to a two-mode cavity setup leads to similar but significantly weaker changes than in the case of coupling to a single cavity mode.
The dipole moment is reduced with increasing coupling strength, see Fig.~\ref{fig:opt_h2o2_2}~a), while the dihedral angle is simultaneously increased, see Fig.~\ref{fig:opt_h2o2_2}~d) orange line.
The other internal coordinates of \ce{H2O2}, shown in Fig.~\ref{fig:opt_h2o_2}~c) and~d), are slightly more affected in the two-mode case, but still in a rather insignificant way.
Similarly to the single-mode coupling, we do not observe any reorientation of the \ce{H2O2} dipole moment with respect to the polarization axes for the two-mode case due to the chosen orientation of the initial geometry.
The dipole moment is parallel to the polarization axis $\bm{e}_1$ and therefore orthogonal to the other, as visualized in Fig.~\ref{fig:opt_h2o2_2}~b).
Next, we discuss the vibro-polaritonic IR spectra of the optimized \ce{H2O2} molecule coupled to a two-mode cavity.
Fig.~\ref{fig:opt_h2o2_2}~e) shows spectra for smaller coupling strengths, which can be roughly divided into two groups.
When $\lambda_m$ is less than \SI{0.05}{\au}, a weak Rabi splitting is observed.
The \gls{lp} transition and the \gls{up} transition are formed by the asymmetric bending mode and the cavity photon mode with the polarization axis $\bm{e}_1$.
This "standard" pair of \gls{lp} and \gls{up} is present in all spectra and the Rabi splitting increases with increasing coupling strength, see Fig.~\ref{fig:opt_h2o2_2}~e) and f).
But similar to the case of \ce{H2O} coupled with two cavity modes, a third weaker signal is present first as a shoulder ($\lambda_m= \SI{0.05}{\au}$ Fig.~\ref{fig:opt_h2o2_2}~e)) later as a distinct peak as shown in Fig.~\ref{fig:opt_h2o2_2}~f).
This weaker "middle" signal is mostly photonic and is characterized by the cavity photon mode with the polarization axis $\bm{e}_2$.
In line with the single-mode case, this transition is due to a weak coupling to the twisting mode in \ce{H2O2} for higher coupling strengths. A visualization of the photonic characters of all relevant transitions can be found in the Supporting Information.

\begin{figure}
     \centering
         \includegraphics[width=0.48\textwidth]{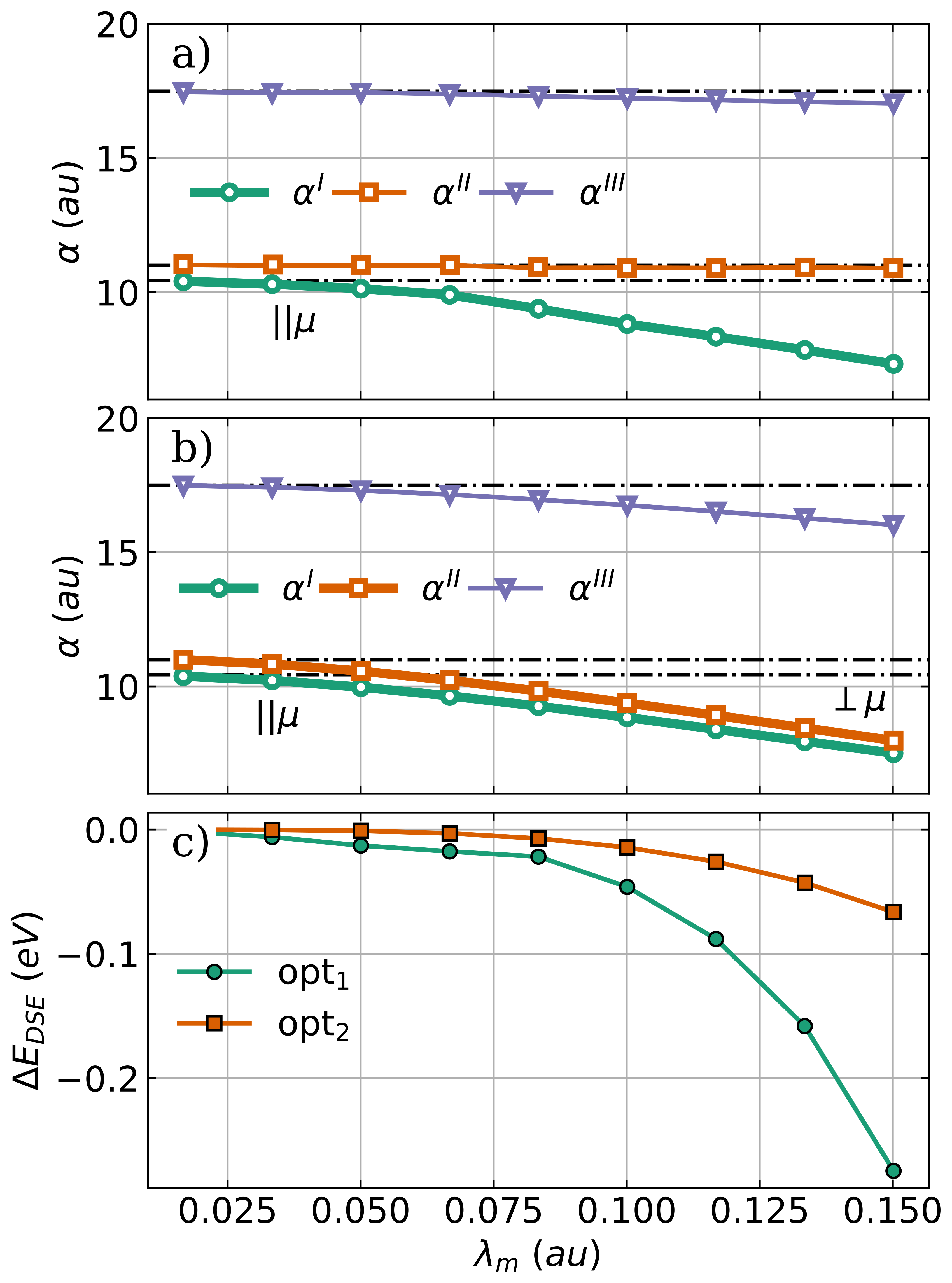}
    \caption{Principal components of the polarizability $\alpha$ as a function of the coupling strength $\lambda_m$ for optimized \ce{H2O2} structures coupled to a single cavity mode a) and to two cavity modes  b). The bold line indicates the component aligned with the cavity polarization axes, and the black dashed dotted lines represent the field-free principal components of the polarizability. c) Change in \gls{dse} contribution as a function of the coupling strength $\lambda_m$  compared to the initial \ce{H2O2} geometry for the single-mode case (green) and the two-mode case (orange). The cavity frequency $\omega_m$ is resonant with the bending mode (\SI{1491}{\per\centi\meter}).}
\label{fig:alpha_h2o2}
\end{figure}

As discussed for the optimization of \ce{H2O}, we also want to understand the observed changes in \ce{H2O2} due to the cavity interaction by analyzing the polarizability $\alpha$ and the \gls{dse} contribution.
The principal components of the polarizability tensor $\alpha$ and the change in the \gls{dse} contribution are shown in Fig.~\ref{fig:alpha_h2o2} as a function of the coupling strength for both optimized \ce{H2O2}-cavity systems.

Consistent with the \ce{H2O} results shown in Fig.~\ref{fig:alpha_h2o}, in both the single-mode case and the two-mode case, all three principal components of the polarizability $\alpha$ decrease with increasing coupling strength, see Fig.~\ref{fig:alpha_h2o2}~a) and b).
The components aligned with the cavity polarization axes, highlighted in bold, are more strongly reduced by the cavity interaction, while especially for the single-mode case Fig.~\ref{fig:alpha_h2o2}~a) the other two components remain almost unchanged.
We do not see reorientation of \ce{H2O2} for the cavity polarization axes studied in the single-mode and two-mode cases, since for the chosen molecular ordination the axes are already aligned with the smallest polarizability components $\alpha^{I}$ and $\alpha^{II}$.
Consequently, the observed reduction in polarizability is achieved only by geometrical changes and the direct response of the electronic structure to the cavity photon field.
These geometrical changes and the electronic response also minimizes the \gls{dse} contribution.
The change in $E_{DSE}$ is plotted as a function of $\lambda_m$ in Fig.~\ref{fig:alpha_h2o2}~c).
Interestingly, the energy change is much larger for the single-mode case (green line) compared to the two-mode case (orange line), although the change in $\alpha$ is comparable.
To explain this, we have to consider that the total \gls{dse} contribution can be separated into a one-electron and two-electron part~\cite{Schnappinger2023-hh,Angelico2023-ll,Sidler2024-vm}. uniquly
The behavior of the one-electron contribution, and usually larger part, can be estimated by the polarizability, while the two-electron contribution is defined by the dipole moment.
Since in the single-mode case \ce{H2O2} planarizes with increasing coupling strength, its dipole moment vanishes and the two-electron part of the \gls{dse} is zero.
Consequently, in the single-mode case, the change in \gls{dse} reflects both the decrease in polarizability and the decrease in the dipole moment.
The present result for the optimization of \ce{H2O} and \ce{H2O2} coupled to an optical cavity clearly shows that not only the dipole moment of the molecule is important, but also the polarizability is a decisive factor in determining the influence of the molecule-cavity interaction.

\section{Conclusion}

To conclude, we studied the importance of geometry relaxation and relative orientation in the context of molecules coupled to the photon modes of optical cavities.
Both effects are currently neglected in most computational studies in the field of polaritonic chemistry.
As two illustrative examples, we have optimized \ce{H2O} and \ce{H2O2} resonantly coupled to one or two cavity photon modes.
For the case of \ce{H2O}, the predominant effect observed during the optimization processes is a rotation that leads to a reorientation of the molecule with respect to the photon modes.
Surprisingly, we could even observe an effect of the restricted rotational degrees of freedom on the vibro-polaritonic IR spectrum for larger coupling strengths.
These results clearly show that rotational motion is no longer an unrestricted degree of freedom in a cavity molecular system that can be neglected.
In contrast, for the optimization of \ce{H2O2}, no rotational motion occurred for the chosen initial conditions (orientation of the molecule and polarization axes of the cavity), and geometric relaxation is more important.
Consequently, the interpretation of computational studies based on fixed molecular structures and fixed orientation should be viewed with caution.
Comparing the optimizations of \ce{H2O2} coupled to one- and two-cavity modes, we want to highlight that the results have some similarities, but also show significant differences.
The more realistic two-mode case with two cavity modes predicts smaller changes in the molecular structure, while the single-mode case overemphasizes them, even leading to a planar structure of \ce{H2O2}.

In agreement with recent studies~\cite{Schnappinger2023-hh,Liebenthal2024-xb}, we can confirm the minimization of the \gls{dse} contribution as the driving force for both orientation and geometrical relaxation.
The \gls{dse} has been shown~\cite{Rokaj2018-ww,Schafer2020-cb,Sidler2022-cg,Schnappinger2023-hh} to be strictly necessary to obtain a finite polarization and a bounded solution.
However, despite its importance, \gls{dse} is still a rather abstract idea, so we derived a more accessible concept to estimate the effect of cavity interaction on molecular geometry.
We are able to explain the observed rotation of the cavity-coupled molecule by determining the main components of the polarizability tensor $\alpha$.
Without fixing its orientation, the molecule will reorient so that the cavity mode polarization axes align with the smallest components of the polarizability.
With respect to geometric relaxation, we observed two effects: First, it aims to reduce the polarizability of the molecule and, second, it also reduces the dipole moment itself.
The evaluation of both the dipole moment and the polarizability is likely to be a useful and simple tool to evaluate the possibility of influencing molecules and chemical reactions by strong light-matter interaction inside an optical cavity.
Exploring the changes in orientation and geometries in molecular ensembles are a logical next step.

\section{Methods}
\subsection{Theoretical background}

Starting from the non-relativistic Pauli-Fierz Hamiltonian in the length gauge representation~\cite{spohn2004dynamics,Ruggenthaler2018-ew,jestadt2019light,lindoy2023quantum,Ruggenthaler2023-ej} we apply the \gls{cboa}~\cite{flick2017atoms,flick2017cavity,Flick2018-ns,Fischer2023-pe} to simulate molecules interacting with the confined electromagnetic field in an optical cavity under \gls{vsc} conditions.
By making use of a generalized Born-Huang expansion~\cite{Schafer2018-vf,Ruggenthaler2023-ej} the cavity modes are grouped with the nuclei and the resulting electronic subsystem can be solved in an \textit{ab-initio} manner~\cite{Sidler2023-vm,Schnappinger2023-hh,Angelico2023-ll}.
Atomic units ($\hbar=4\pi\varepsilon_0=m_e=1$) are used in the following unless otherwise noted, and bold symbols denote vectors.
The electronic \gls{cboa} Hamiltonian for multiple cavity modes takes the form of
\begin{equation}
\label{eq:h_cbo}
\hat{H}_{CBO} = \hat{H}_{el} +  \sum_{m}^{N_{M}} \frac{1}{2} \omega_{m}^2 q_{m}^2 - \omega_{m} q_{m} \left(\bm{\lambda}_{m} \cdot \bm{\hat{\mu}} \right) + \frac{1}{2} \left(\bm{\lambda}_{m} \cdot \bm{\hat{\mu}} \right)^2\,,
\end{equation}
where $\bm{\hat{\mu}}$ represents the molecular dipole operator, which is defined by the operators of the $N_{el}$ electron coordinates $\bm{\hat{r}}$ and the classic coordinates $\bm{R}$ of the $N_{Nuc}$ nuclei.
$\hat{H}_{el}$ is the Hamiltonian for the field-free many-electron system.
Each of the $N_m$ cavity modes contributes three additional terms to the electronic Hamiltonian. The first term is a harmonic potential introduced by the photon displacement field, with the classic photon displacement coordinate $q_{m}$ and $\omega_m$ being the frequency of the cavity mode.
The second term describes the dipole coupling between the molecular system and the photon displacement field, which is characterized by the coupling strength $\bm{\lambda}_{m}$.
The last term is the \gls{dse} operator~\cite{Rokaj2018-ww,Schafer2020-cb,Sidler2022-cg}, which is an energy contribution that describes the self-polarization of the molecule-cavity system.
The cavity mode-specific coupling parameter $\bm{\lambda}_{m}$ for a cavity with effective mode volume $V_m$ is defined as follows:
 \begin{equation}
 \label{eq:lam}
\bm{\lambda}_{m} =  \bm{e}_{m} \lambda_{m} =  \bm{e}_{m}  \sqrt{\frac{4 \pi}{V_m}}\,.
\end{equation}
The unit vector $\bm{e}_{m}$ denotes the polarization axis of the cavity mode $m$.

The many-electron problem described by Eq.~\ref{eq:h_cbo} can be solved using the \gls{cbohf} approach~\cite{Schnappinger2023-hh}. The resulting energy $E_{CBO}$ is a function of the nuclear coordinates $\bm{R}$ and the photon displacement coordinates $\bm{q}$, represented as a vector grouping all $q_{m}$:
\begin{align}
E_{CBO} = &\bigl\langle  E_{CBO} \bigr\rangle (\bm{R},\bm{q})  = \bigl\langle E_{el} \bigr\rangle (\bm{R}) \\ &+ \sum_{m}^{N_{M}}  \bigl\langle E_{lin}^{({m})} \bigr\rangle (\bm{R},q_{m}) + \bigl\langle E_{dse}^{({m})} \bigr\rangle (\bm{R}) +E_{dis}^{({m})} (q_{m})\nonumber
\end{align}
The first derivative of the energy $E_{CBO}$ with respect to the nuclear and photon displacement coordinates can be calculated analytically~\cite{Schnappinger2023-wp} and defines the \gls{cbohf} gradient $\bm{g}_{CBO}$ as a $\left( 3N_{A} +N_{M} \right)$ vector, where $N_{A}$ is the number of atoms in the molecule.
\begin{equation}
\boldsymbol{g} =\bm{\nabla} E_{CBO}
\end{equation}
Based on the analytic gradients, the CBO-Hessian matrix $\boldsymbol{H}$ of size $\left(3N_{A} +N_{M}\right)\left(3N_{A} +N_{M}\right)$ is accessible via finite differences~\cite{Schnappinger2023-wp}.
\begin{equation}
\boldsymbol{H}_{ij} = \bm{\nabla}_{i} \boldsymbol{g}_{j} \approx \frac{\boldsymbol{g}_j\left(x_{i}+\Delta \right)-\boldsymbol{g}_j\left(x_{i}-\Delta \right)}{2\Delta}
\end{equation}
with $x_{i}$ being a nuclear or displacement coordinate.
The molecular polarizability tensor $\alpha$ at the \gls{cbohf} level is calculated as the first derivative of the \gls{cbohf} dipole moment vector with respect to a small external field, also using finite differences.
Both the analytic gradient and the numerical Hessian can be used to optimize molecules coupled to cavity photon modes.
In this manuscript we use the \gls{bf} algorithm:
\begin{equation}
    \boldsymbol{x}_{n+1} = \boldsymbol{x}_{n} - \boldsymbol{\tilde{H}}^{-1} \bm{g}_{CBO}\,,
\end{equation}
where $\boldsymbol{x}$ is the combined nuclear and displacement coordinate vector $\left( 3N_{A} +N_{M} \right)$ and $\bm{g}_{CBO}$ the gradient vector.
The approximate Hessian matrix $\boldsymbol{\tilde{H}}$ at each point $n$ is updated with the Hessian matrix at stage $n-1$ according to:
\begin{equation}
 \boldsymbol{\tilde{H}}_{n} = \boldsymbol{\tilde{H}}_{n-1} + \frac{\Delta \boldsymbol{ g} \cdot  \Delta \boldsymbol{g}^T}{\Delta \boldsymbol{g}^T \cdot  \Delta \boldsymbol{ x}} - \frac{\boldsymbol{\tilde{H}}_{n-1} \cdot \Delta \boldsymbol{x} \cdot \Delta \boldsymbol{x}^T \boldsymbol{\tilde{H}}_{n-1}}{\Delta \boldsymbol{x}^T \cdot \boldsymbol{\tilde{H}}_{n-1} \cdot \Delta \boldsymbol{x}}
\end{equation}
A detailed benchmark of the implemented \gls{bf} algorithm against the Steepest Descent method and the Newton–Raphson method can be found in the Supporting Information.

\subsection{Computational details}

The \gls{bf} algorithm, the necessary analytical gradients and the numerical Hessian for the \gls{cbohf} ansatz have been implemented in the Psi4NumPy environment~\cite{Smith2018-tu}, which is an extension of the PSI4~\cite{Smith2020-kq} electronic structure package.
All calculations were performed using the aug-cc-pVDZ basis set~\cite{Kendall1992-wu} and all geometries were pre-optimized at the Hartree-Fock level of theory.
In all \gls{cbohf} calculations performed in this work, we consider a lossless cavity.
The coupling strength $\lambda_m$ is chosen between \SI{0.015}{\au} and \SI{0.150}{\au} to assess the medium and strong coupling situation in the case of a single molecule.
We emphasize that parts of the light-matter coupling used here are significantly larger than what can presently be achieved in experiments.
For example, $\lambda_m = \SI{0.1}{\au}$ corresponds to an effective mode volume of less than \SI{0.2}{\nano\meter\cubed} , which is less than the typical mode volumes of approximately \SI{10.0}{\nano\meter\cubed} that can be achieved in plasmonic cavities~\cite{Benz2016-vq,Mondal2022-le}.

The \gls{cbohf} energy and gradients depend on the relative orientation of the molecule and the polarization axes of the cavity.
For this reason, the internal coordinate system traditionally used to optimize molecular geometries is a poor choice because it neglects spatial orientation.
Therefore, all geometry optimizations in this work were performed using Cartesian coordinates.
For the first step of the \gls{bf} optimization, the exact Hessian matrix was computed, and to improve convergence for the \ce{H2O} optimization, the exact Hessian was recomputed after every 10th step.
The maximum component and the root mean square deviation of the gradient and the displacement vector are used as convergence criteria.
For the gradient, both must be less than \SI{e-5}{\au} and for the displacement vector, both must be less than \SI{e-4}{\au}.
The semiclassical harmonic approximation~\cite{Schnappinger2023-wp} was used to determine the normal modes and frequencies of the optimized coupled cavity-molecular systems.
All optimized structures discussed in this manuscript are real minima, since they have no imaginary frequencies.

\begin{acknowledgments}
M.K. acknowledges funding from the European Union’s Horizon 2020 research.
This project has received funding from the European Research Council (ERC) under the European Union’s Horizon 2020 research and innovation program (grant agreement no. 852286).
\end{acknowledgments}

\section*{Author Contributions}
T.S. and M.K. designed the research. T.S. implemented the optimization routines and performed the simulations. T.S. and M.K contributed equally on analyzing the data and writing the manuscript.

\section*{Competing interests}
The authors declare no competing interests.

\section*{Supplementary information}
The supplementary material includes a detailed analysis of vibro-polaritonic normal modes, a benchmark optimization method and parameters, and all optimized geometries. Correspondence and requests for materials should be addressed to T.S. or M.K.

\bibliography{lit.bib}

\end{document}


\maketitle

\tableofcontents
\clearpage

\section{Vibro-Polaritonic
Normal Modes Analysis}

The general concept of performing a normal mode analysis in the \gls{cboa} was introduced in our previous work~\cite{Schnappinger2023-wp}, and the reader is referred to the paper for details of the theory. 
For convenience, the main ideas are summarized in the following.
The harmonic approximation gives access to the normal modes $\bm{a}_i$. 
In \gls{cboa} the normal mode vectors have terms $a_c$ describing the change in the classical photon displacement field coordinates $q_m$. 
The value of $|a_c|^{2}$ for a given normal mode is a measure of how strongly the corresponding vibrational transition interacts with the photon field. 
For an uncoupled light-matter system, a pure molecular transition is characterized by a $|a_c|^{2}$ value of zero, whereas the bare photon mode has a value of one. 
Note that due to the length gauge description $q_m$ and the corresponding value $a_c$ are no longer a pure photonic quantity if light and matter are coupled~\cite{Rokaj2018-ww,Schafer2020-cb,Welakuh2023-ap}. 
However, $|a_c|^{2}$ can still be used as a probe to identify how photonic the corresponding vibrational transition is. The information obtained is comparable to the coefficients in the Hopfield models~\cite{Hopfield1958-ms}.

In the following, we use the $|a_c|^{2}$ values to characterize all relevant transitions in the vibro-polaritonic IR spectra of all optimized structures.
For the case of the coupled \ce{H2O}-cavity system, we also discuss the signal intensities $\mathcal{I}$ in the harmonic approximation.
The intensities are calculated as the projection of the dipole moment gradient on the normal mode vectors $\bm{a}_i$:
\begin{equation}
\mathcal{I}_i  = \left( \bm{\nabla} \bigl\langle \bm{\hat{\mu}} \bigr\rangle \cdot \bm{a}_i \right)^2\,.
\end{equation}

\begin{figure}[htb!]
     \centering         \includegraphics[width=0.7\textwidth]{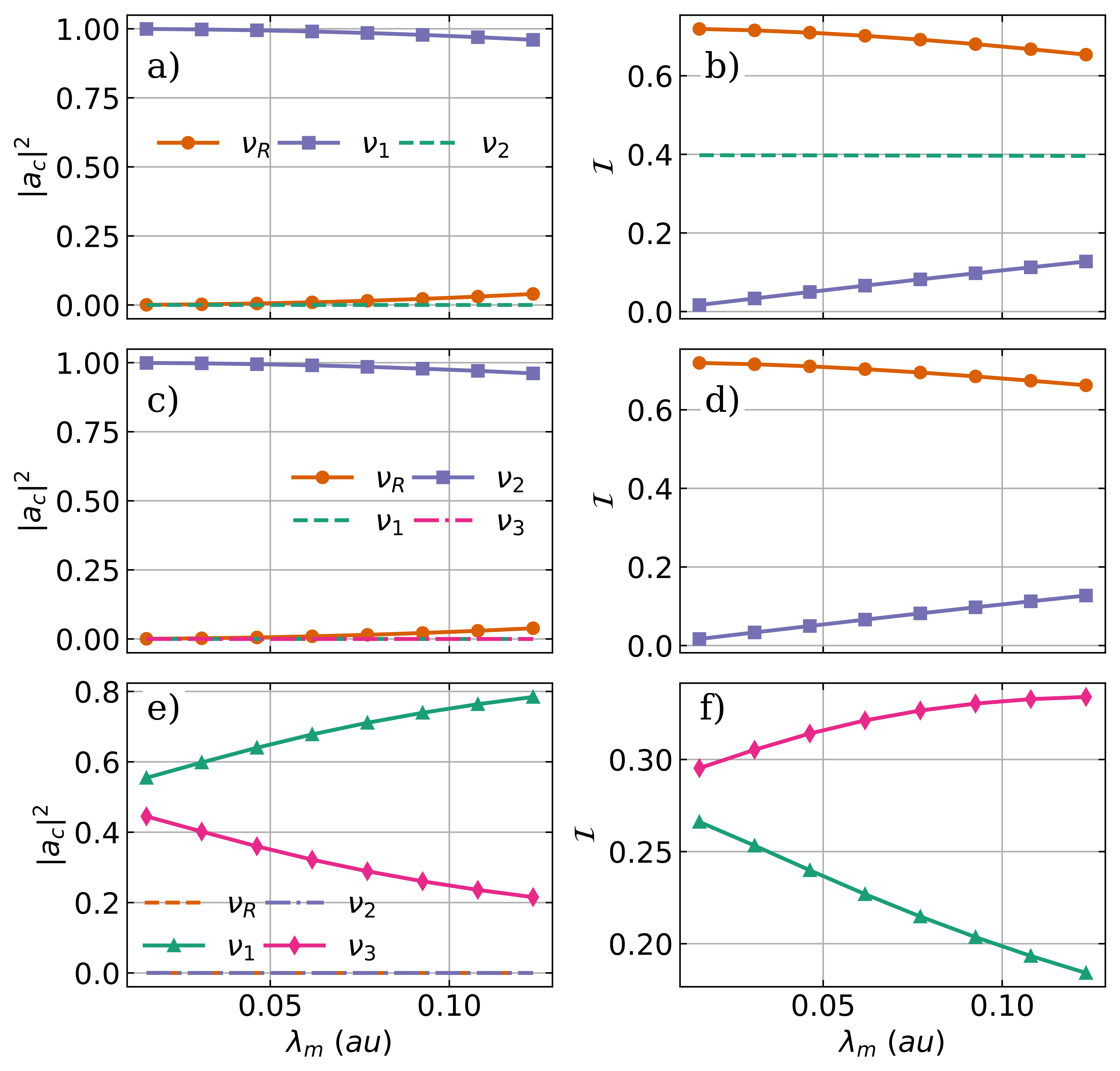}
    \caption{Relevant $|a_c|^{2}$ values and harmonic intensities $\mathcal{I}$ as a function of the coupling strength $\lambda_m$ for optimized \ce{H2O} structures coupled to a single cavity mode a) and b) as well as to two orthogonal cavity modes c), d), e) and f). The values for the cavity mode with the polarization axis $\bm{e}_2$ orthogonal to the dipole moment are given in c) and d) and for the one with the polarization axis $\bm{e}_1$ parallel to the dipole moment in e) and f). The cavity frequency $\omega_c$ is \SI{1744}{\per\centi\meter} and the cavity coupling $\lambda_{m}$ increases from \SI{0.015}{\au} to \SI{0.123}{\au}.}
\label{fig:h2o_coef}
\end{figure}

Fig.~\ref{fig:h2o_coef} shows the $|a_c|^{2}$ values and the harmonic intensities $\mathcal{I}$ of the relevant transitions as a function of the coupling strength $\lambda_m$ for optimized \ce{H2O}-cavity systems.

In the case of a single cavity mode coupled to \ce{H2O}, only three transitions are relevant to explain the spectral features in the vicinity of \SI{1744}{\per\centi\meter}. 
The main transition $\nu_2$ shown as a green dashed line in Fig.~\ref{fig:h2o_coef} a) and b) corresponds to the molecular bending mode. 
It does not hybridize with the photon mode and its intensity is constant with increasing $\lambda_m$.
In contrast, $\nu_1$, shown as purple in Fig.~\ref{fig:h2o_coef} a) and b), is purely photonic at low coupling and has an intensity close to zero.
With increasing coupling strength, $\nu_1$ begins to weakly pair to a rotational mode ($\nu_R$ orange line).
As a consequence, $\nu_R$ becomes slightly photonic and at the same time $\nu_1$ gains intensity.

In the case of \ce{H2O} coupled to two orthogonal cavity modes, we divide the discussion into two parts: First, the photon mode orthogonal to the dipole moment shown in Fig.~\ref{fig:h2o_coef} c) and d) and second, the one parallel to the dipole mode shown in Fig.~\ref{fig:h2o_coef} e) and f).
For the orthogonal cavity mode, the picture is almost identical to the single-mode case. 
The photon mode transition $\nu_2$ starts almost decoupled and dark but starts to weakly couple to the rotational mode transition $\nu_R$ with increasing coupling strength. 
In contrast, in the case of the parallel cavity mode, the transitions $\nu_1$ and $\nu_3$ are clearly hybridized already for low coupling strengths and stay so for increasing $\lambda_m$.

\begin{figure}[htb!]
     \centering
\includegraphics[width=0.5\textwidth]{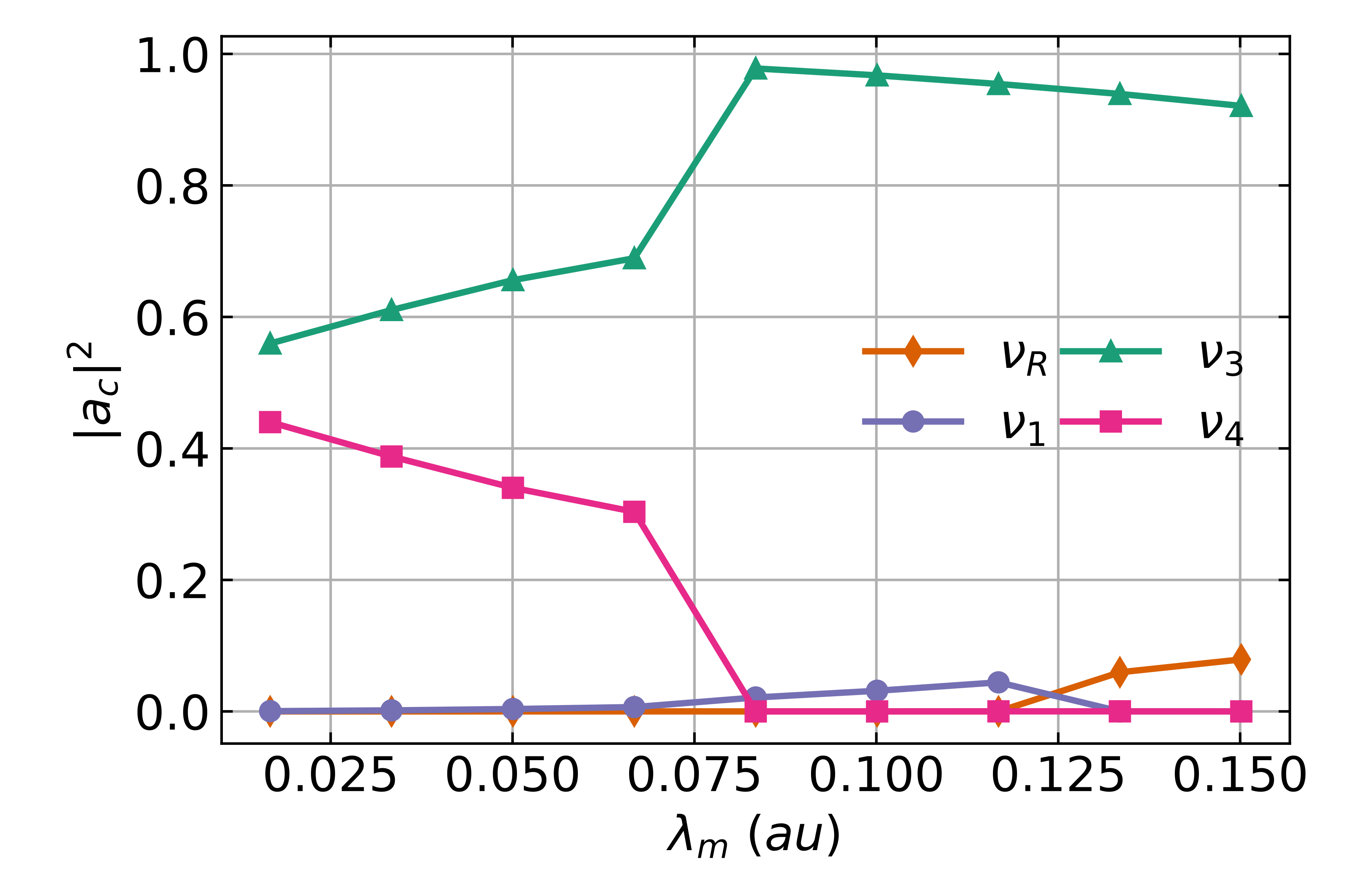}
    \caption{$|a_c|^{2}$ values of the four relevant transitions for the optimized \ce{H2O2} molecule coupled to a single cavity mode as a function of the coupling strength $\lambda_m$.  The cavity frequency $\omega_c$ is \SI{1491}{\per\centi\meter} and the cavity coupling $\lambda_{m}$ increases from \SI{0.015}{\au} to \SI{0.150}{\au}.}
\label{fig:h2o2_m1_coef}
\end{figure}

In Fig.~\ref{fig:h2o2_m1_coef} we show the $|a_c|^{2}$ values of the four relevant transitions for the optimized \ce{H2O2} molecule coupled to a single cavity mode as a function of the coupling strength $\lambda_m$.
The asymmetric bending mode transition $\nu_4$ (pink) and the cavity photon mode transition $\nu_3$ (green) are clearly hybridized for low coupling strengths, smaller than \SI{0.075}{\au}. 
Due to geometrical changes, the cavity photon mode transition $\nu_3$ decouples from $\nu_4$ for larger $\lambda_m$ and starts to interact weakly with the \ce{H2O2} twisting mode ($\nu_1$ purple). For even higher coupling strengths and close to planarization, $\nu_3$ is weakly coupled to a rotational mode ($\nu_R$ orange), similar to the \ce{H2O} case.

\begin{figure}[ht!]
     \centering    \includegraphics[width=0.74\textwidth]{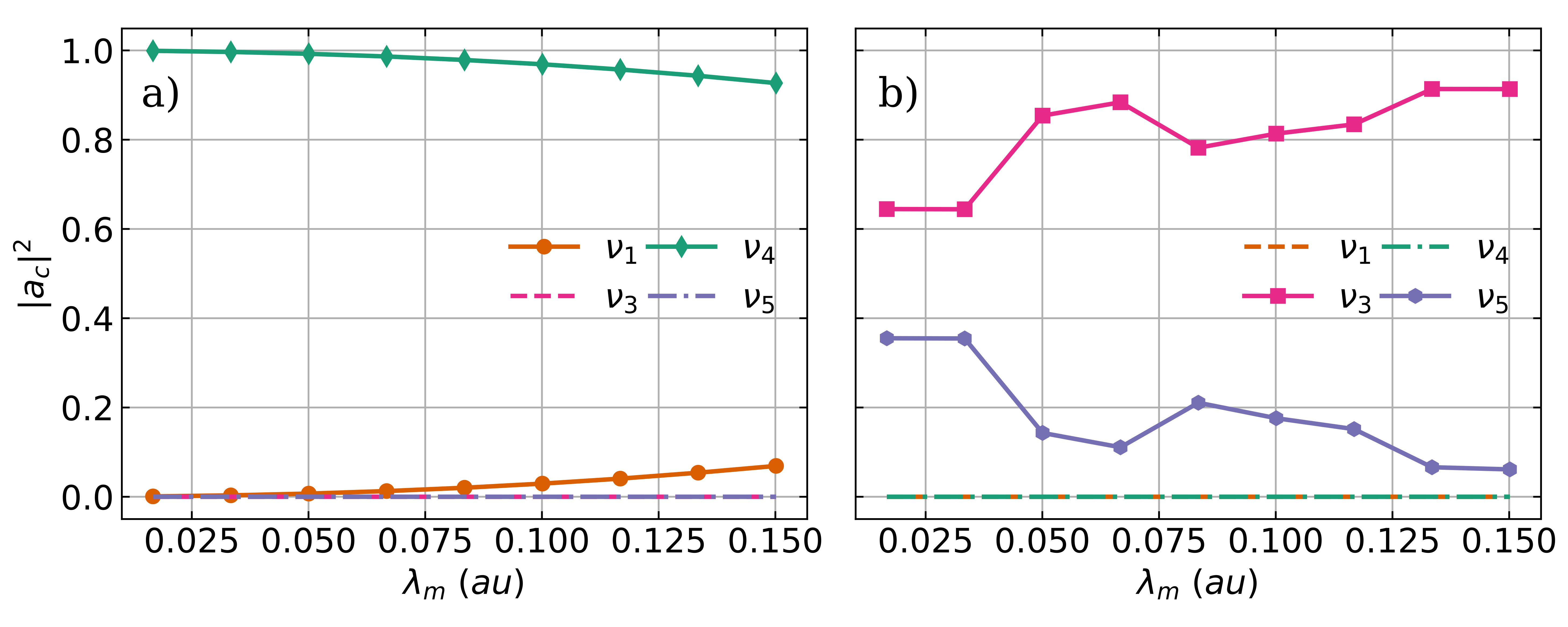}
    \caption{$|a_c|^{2}$ values of the four relevant transitions for the optimized \ce{H2O2} molecule coupled to two orthogonal cavity modes as a function of the coupling strength $\lambda_m$. a) For the cavity mode with the polarization axis $\bm{e}_2$ orthogonal to the dipole moment and b) for the one with the polarization axis $\bm{e}_1$ parallel to the dipole moment.
    The cavity frequency $\omega_c$ is \SI{1491}{\per\centi\meter} and the cavity coupling $\lambda_{m}$ increases from \SI{0.015}{\au} to \SI{0.150}{\au}.}
\label{fig:h2O2_m2_coef}
\end{figure}

The $|a_c|^{2}$ values of the four relevant transitions for the optimized \ce{H2O2} molecule coupled to two orthogonal cavity modes are depicted in Fig.~\ref{fig:h2O2_m2_coef}. 

For the orthogonal cavity mode Fig.~\ref{fig:h2O2_m2_coef}~a), 
the corresponding photon mode transition $\nu_4$ (green) starts almost decoupled and dark, but starts to couple weakly to the \ce{H2O2} twisting mode ($\nu_1$ orange) with increasing coupling strength. 
In contrast, in the case of the parallel cavity mode Fig.~\ref{fig:h2O2_m2_coef}~b), the transitions $\nu_3$ and $\nu_5$ are clearly hybridized already at low coupling strengths and stay so for increasing $\lambda_m$.

\section{Benchmark Optimization Methods and Parameters}

We benchmark the performance of various algorithms for the optimization of coupled molecular-cavity systems in the \gls{cboa} representation. 
As a test case, we optimize a \ce{H2O} molecule coupled to a cavity. For the case of a single cavity mode, $\bm{e}$ is neither parallel nor orthogonal to the molecular plane. 
In the case of two modes, $\bm{e}_1$ is in the molecular plane and $\bm{e}_2$ is aligned with the normal vector of the molecular plane.
The simplest algorithm used is the \gls{sd} method, which requires only the gradient in each optimization step. 
The most computationally intensive is the \gls{nr} method, which requires the exact Hessian matrix in each step.
The \gls{bf} algorithm offers a middle ground between \gls{sd} and \gls{nr} in terms of computational complexity.
It uses an updating scheme that approximates the Hessian matrix at each point $n$ using the gradient, the displacement, and the Hessian matrix at step $n-1$.
For the simplest version of the \gls{bf} method, the identity matrix is used as an approximate Hessian at $n=0$. 
To improve the performance, the exact Hessian can be used in the first step and during the optimization the exact Hessian can be recomputed after a certain number of steps.
We have included an augmented Hessian method for each optimization routine in which we use the exact Hessian to ensure that the Hessian matrix is positively defined.
\begin{equation}
    \boldsymbol{x}_{n+1} = \boldsymbol{x}_{n} - \left( \boldsymbol{H} - \left( \min \left(\epsilon\right) + \gamma \right))\boldsymbol{1}\right)^{-1} \boldsymbol{g}\,.
\end{equation}
Here, $\min \left(\epsilon\right)$ is the smallest eigenvalue of the Hessian matrix and $\gamma$ is a small number to prevent the augmented Hessian matrix from being non-invertible.
The augmentation is necessary to avoid inadvertently optimizing a transition state or a higher-order saddle point. 

\begin{figure}[htb!]
     \centering     \includegraphics[width=0.75\textwidth]{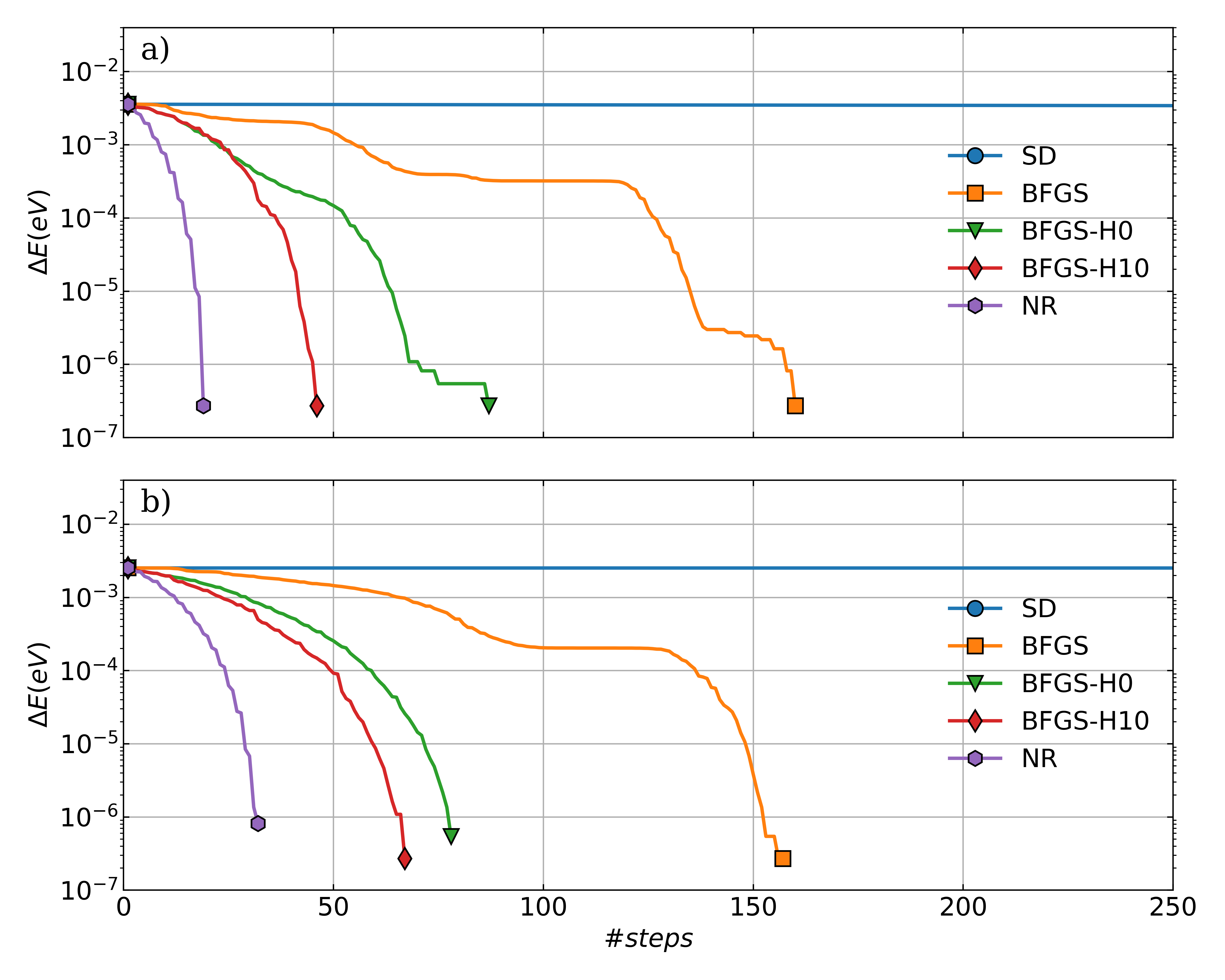}
    \caption{Energy convergence for the optimization of a single \ce{H2O} molecule coupled to a) one cavity photon mode or b) two cavity photon modes for $\omega_m = \SI{1744.0}{\per\centi\meter}$ and $\lambda_m = \SI{0.031}{\au}$ using different optimization routines. The different optimization routines are color-coded. The different versions of the \gls{bf} method are labeled as follows: BFGS started with the identity matrix and no Hessian was computed, BFGS-H0 started with the exact Hessian, and BFGS-H10 started with the exact Hessian and recomputed it every 10th step. The starting point and the converged end point are marked.}
\label{fig:bench_method}
\end{figure}

Fig.~\ref{fig:bench_method} shows the convergence behavior for different optimization algorithms.

\begin{figure}[htb!]
     \centering
         \includegraphics[width=0.75\textwidth]{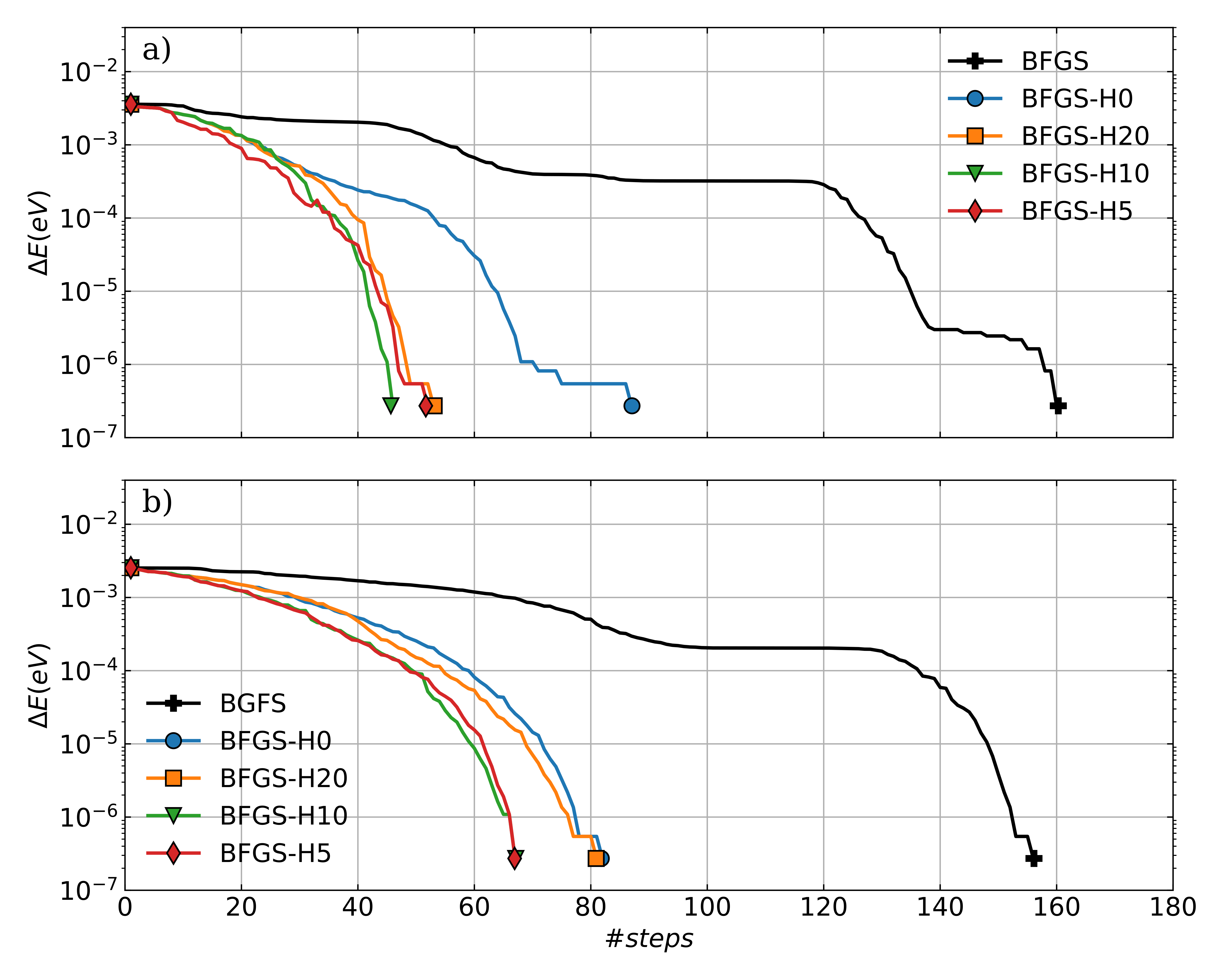}
    \caption{Energy convergence for the optimization of a single \ce{H2O} molecule coupled to a) one cavity photon mode or b) two cavity photon modes for $\omega_m = \SI{1744.0}{\per\centi\meter}$ and $\lambda_m = \SI{0.031}{\au}$ using different versions of the \gls{bf} routine. Different intervals for recalculation of the exact Hessian are color-coded. The\gls{bf} results without the exact Hessian are shown in black.  The starting point and the converged end point are marked.}
\label{fig:bench_hx}
\end{figure}

The \gls{sd} method does not converge within the maximum number of 250 steps. 
The \gls{nr} method and all versions of the \gls{bf} method converge to the same minimum structure. 
As expected, the \gls{nr} method shows the fastest convergence, while the simplest \gls{bf} version needs about three times more optimization steps. 
Including Hessian information in the \gls{bf} algorithm significantly speeds up convergence, especially when the exact Hessian is recomputed during optimization.
In Fig.~\ref{fig:bench_hx} the energy convergence for different intervals for the reactivation of the Hessian is shown.

\begin{figure}[htb!]
     \centering
         \includegraphics[width=0.75\textwidth]{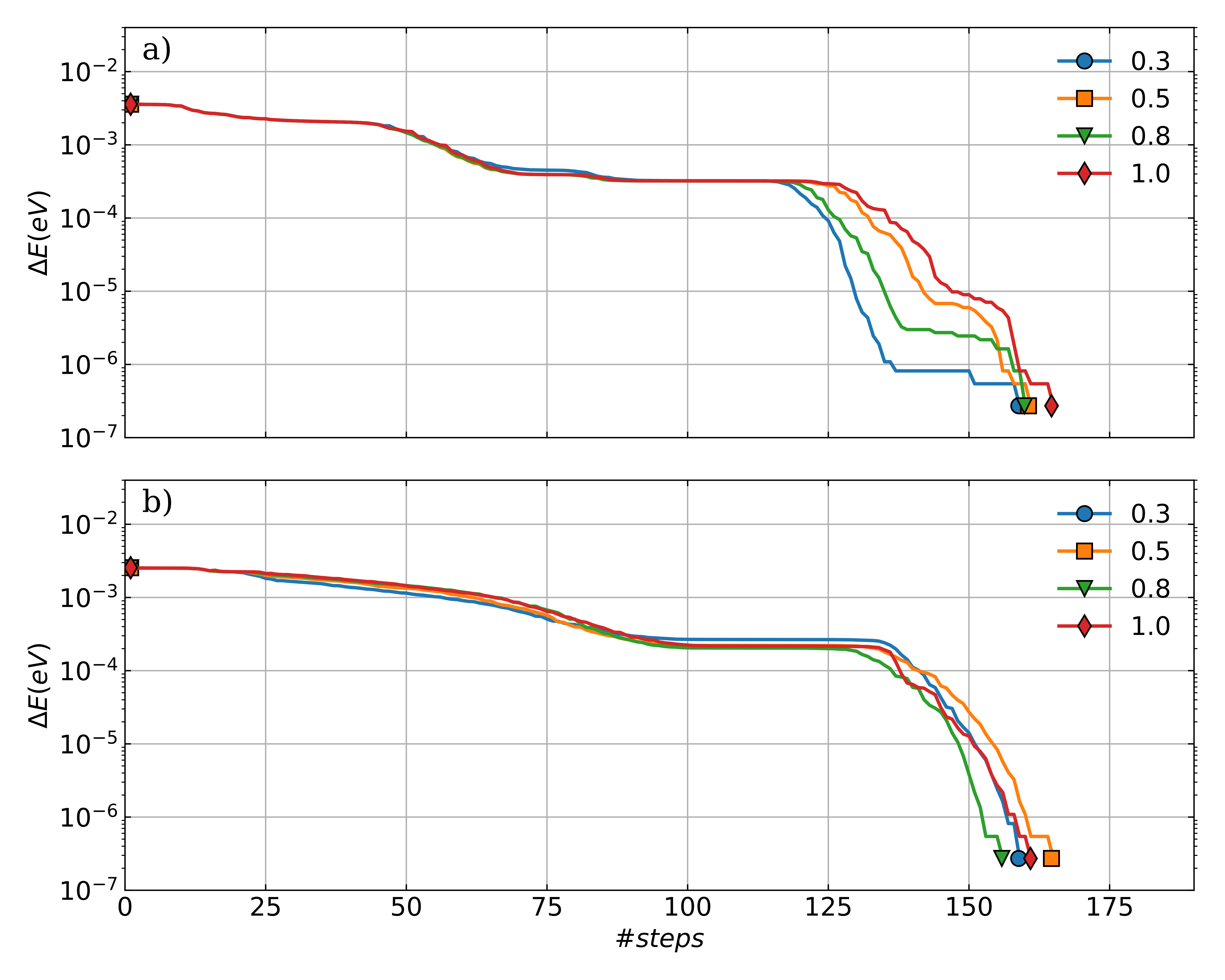}
    \caption{Energy convergence for the optimization of a single \ce{H2O} molecule coupled to a) one cavity photon mode or b) two cavity photon modes for $\omega_m = \SI{1744.0}{\per\centi\meter}$ and $\lambda_m = \SI{0.031}{\au}$ using the \gls{bf} routine recalculating the Hessian in every 10th step and 5 steps in the backtracking line search. Different values for the trust radius $R$ are color-coded. The starting point and the converged end point are marked.}
\label{fig:bench_tr}
\end{figure}

For the case of \ce{H2O} coupled to a single cavity mode (Fig.~\ref{fig:bench_tr}~a)), recalculating the exact Hessian during optimization significantly speeds up convergence.
However, changing the interval from every 20th step to every 5th step shows only a small improvement. 
This finding also holds for the case of two orthogonal cavity modes, shown in Fig.~\ref{fig:bench_tr}~b).

\begin{figure}[htb!]
     \centering
         \includegraphics[width=0.75\textwidth]{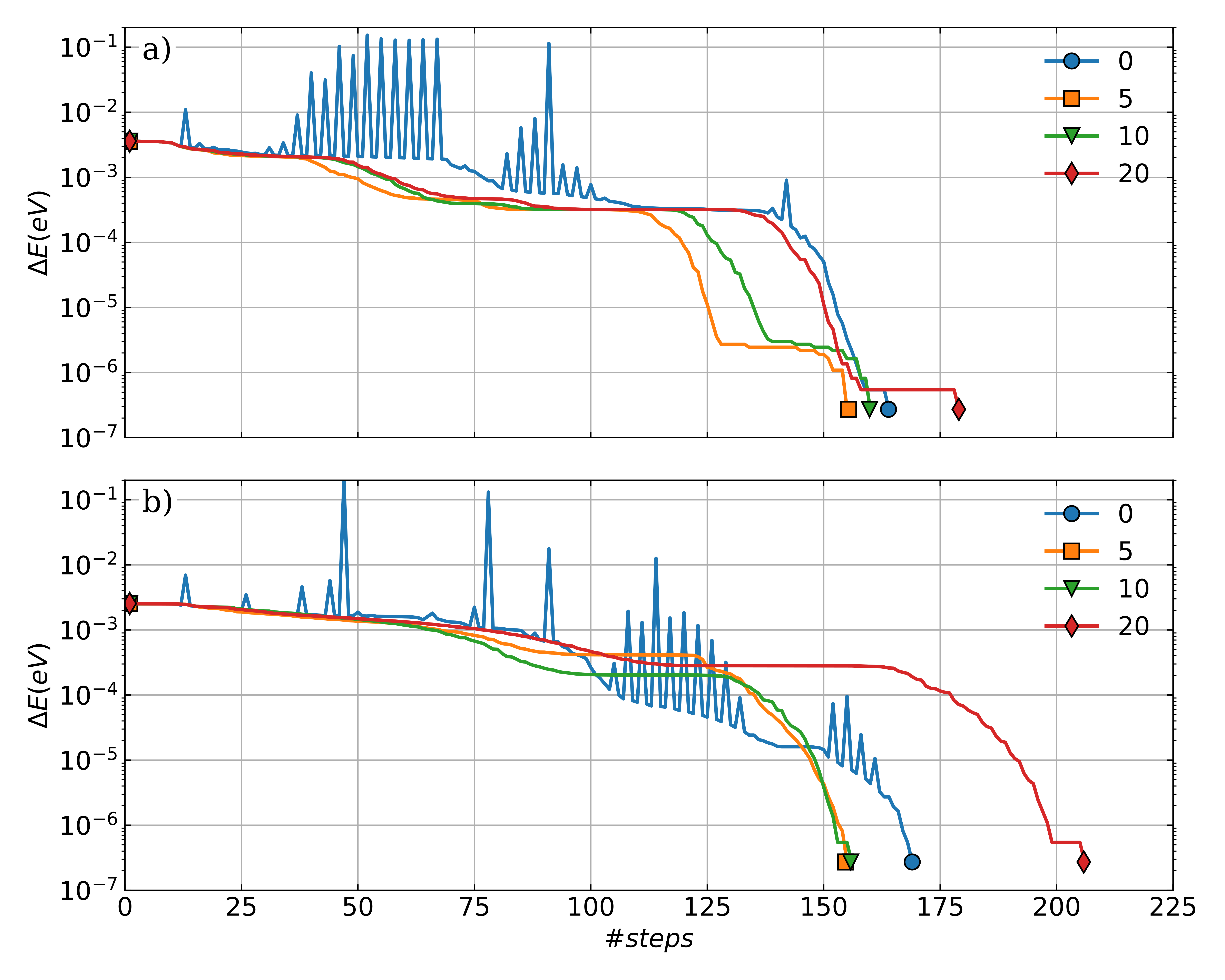}
    \caption{Energy convergence for the optimization of a single \ce{H2O} molecule coupled to a) one cavity photon mode or b) two cavity photon modes for $\omega_m = \SI{1744.0}{\per\centi\meter}$ and $\lambda_m = \SI{0.031}{\au}$ using the \gls{bf} routine recalculating the Hessian in every 10th step and a trust radius of \SI{0.8}{\au}. Different numbers of steps in the backtracking line search are color-coded. The starting point and the converged end point are marked.}
\label{fig:bench_ls}
\end{figure}

To improve convergence and overall stability, we implement a trust-radius approach combined with a backtracking line search in our optimization routines. 
The trust-radius approach sets an upper bound $R$ for the step size, while the backtracking line search further reduces the step size if the energy increases along the step.
The latter comes with a coast of additional single-point calculations during each optimization step. 
The energy convergence for different trust radii $R$ is shown in Fig.~\ref{fig:bench_tr} and for different number of iterations of the backtracking line search is shown in Fig.~\ref{fig:bench_ls}.

The chosen size of the trust radius $R$ has only a small influence on convergence, and a value of \SI{0.8}{\au} shows the best results (green line in Fig.~\ref{fig:bench_tr}). 
In contrast, the number of iterations of the backtracking line search has an impact on both stability and convergence.
With more steps in the line search, the average step size is getting smaller, and therefore more steps are needed. 
However, without line search (blue line in Fig.~\ref{fig:bench_ls}), the energy change is highly fluctuating, making optimization more unstable.
The best results in both stability and convergence are achieved for 5 steps in the backtracking line search (orange line in Fig.~\ref{fig:bench_ls}) 

\clearpage

\section{Optimized geometries}

\subsection{Optimized \ce{H2O} coupled to a single cavity mode}

All optimizations were performed using the \gls{bf} method, starting from the exact Hessian matrix in the first step and recalculating it after every 10th step.
The minima found were verified to have no imaginary frequency.

\begin{table}
  \caption{$\lambda_m = \SI{0.015}{\au}$ and $q_1 = \SI{0.0000}{\au}$.}
    \centering
    \begin{tabular}{ c ccc}       
    \hline
      O  & \tablenum[table-format=-1.4]{0.0000} &  \tablenum[table-format=-1.4]{-0.0282} &  \tablenum[table-format=-1.4]{-0.1101} \\
      H  & \tablenum[table-format=-1.4]{0.0000} &  \tablenum[table-format=-1.4]{-0.6170} &  \tablenum[table-format=-1.4]{0.6273} \\
      H  & \tablenum[table-format=-1.4]{0.0000} &  \tablenum[table-format=-1.4]{0.8425} &  \tablenum[table-format=-1.4]{0.2536}    
    \end{tabular}
\end{table}

\begin{table}
  \caption{$\lambda_m = \SI{0.031}{\au}$ and $q_1 = \SI{0.0000}{\au}$.}
    \centering
    \begin{tabular}{ c ccc}       
    \hline
      O  & \tablenum[table-format=-1.4]{0.0000} &  \tablenum[table-format=-1.4]{-0.0260} &  \tablenum[table-format=-1.4]{-0.1107} \\
      H  & \tablenum[table-format=-1.4]{0.0000} &  \tablenum[table-format=-1.4]{-0.6294} &  \tablenum[table-format=-1.4]{0.6146} \\
      H  & \tablenum[table-format=-1.4]{0.0000} &  \tablenum[table-format=-1.4]{0.8370} &  \tablenum[table-format=-1.4]{-0.2707}    
    \end{tabular}
\end{table}

\begin{table}
  \caption{$\lambda_m = \SI{0.046}{\au}$ and $q_1 = \SI{0.0000}{\au}$.}
    \centering
    \begin{tabular}{ c ccc}       
    \hline
      O  & \tablenum[table-format=-1.4]{0.0000} &  \tablenum[table-format=-1.4]{-0.0252} &  \tablenum[table-format=-1.4]{-0.1108} \\
      H  & \tablenum[table-format=-1.4]{0.0000} &  \tablenum[table-format=-1.4]{-0.6335} &  \tablenum[table-format=-1.4]{0.6107} \\
      H  & \tablenum[table-format=-1.4]{0.0000} &  \tablenum[table-format=-1.4]{0.8349} &  \tablenum[table-format=-1.4]{0.2766}    
    \end{tabular}
\end{table}

\begin{table}
  \caption{$\lambda_m = \SI{0.062}{\au}$ and $q_1 = \SI{0.0000}{\au}$.}
    \centering
    \begin{tabular}{ c ccc}       
    \hline
      O  & \tablenum[table-format=-1.4]{0.0000} &  \tablenum[table-format=-1.4]{-0.0070} &  \tablenum[table-format=-1.4]{-0.1134} \\
      H  & \tablenum[table-format=-1.4]{0.0000} &  \tablenum[table-format=-1.4]{-0.7233} &  \tablenum[table-format=-1.4]{0.5000} \\
      H  & \tablenum[table-format=-1.4]{0.0000} &  \tablenum[table-format=-1.4]{0.7791} &  \tablenum[table-format=-1.4]{0.4075}    
    \end{tabular}
\end{table}

\begin{table}
  \caption{$\lambda_m = \SI{0.077}{\au}$ and $q_1 = \SI{0.0000}{\au}$.}
    \centering
    \begin{tabular}{ c ccc}       
    \hline
      O  & \tablenum[table-format=-1.4]{0.0000} &  \tablenum[table-format=-1.4]{-0.2666} &  \tablenum[table-format=-1.4]{-0.1105} \\
      H  & \tablenum[table-format=-1.4]{0.0000} &  \tablenum[table-format=-1.4]{-0.6247} &  \tablenum[table-format=-1.4]{0.6182} \\
      H  & \tablenum[table-format=-1.4]{0.0000} &  \tablenum[table-format=-1.4]{0.8379} &  \tablenum[table-format=-1.4]{0.2655}    
    \end{tabular}
\end{table}

\begin{table}
  \caption{$\lambda_m = \SI{0.093}{\au}$ and $q_1 = \SI{0.0000}{\au}$.}
    \centering
    \begin{tabular}{ c ccc}       
    \hline
      O  & \tablenum[table-format=-1.4]{0.0000} &  \tablenum[table-format=-1.4]{-0.0036} &  \tablenum[table-format=-1.4]{0.1136} \\
      H  & \tablenum[table-format=-1.4]{0.0000} &  \tablenum[table-format=-1.4]{-0.7372} &  \tablenum[table-format=-1.4]{-0.4779} \\
      H  & \tablenum[table-format=-1.4]{0.0000} &  \tablenum[table-format=-1.4]{0.7658} &  \tablenum[table-format=-1.4]{-0.4305}    
    \end{tabular}
\end{table}

\begin{table}
  \caption{$\lambda_m = \SI{0.101}{\au}$ and $q_1 = \SI{0.0000}{\au}$.}
    \centering
    \begin{tabular}{ c ccc}       
    \hline
      O  & \tablenum[table-format=-1.4]{0.0000} &  \tablenum[table-format=-1.4]{0.0099} &  \tablenum[table-format=-1.4]{0.1132} \\
      H  & \tablenum[table-format=-1.4]{0.0000} &  \tablenum[table-format=-1.4]{-0.7880} &  \tablenum[table-format=-1.4]{-0.3874} \\
      H  & \tablenum[table-format=-1.4]{0.0000} &  \tablenum[table-format=-1.4]{0.7091} &  \tablenum[table-format=-1.4]{-0.5179}    
    \end{tabular}
\end{table}

\begin{table}
  \caption{$\lambda_m = \SI{0.123}{\au}$ and $q_1 = \SI{0.0000}{\au}$.}
    \centering
    \begin{tabular}{ c ccc}       
    \hline
      O  & \tablenum[table-format=-1.4]{0.0000} &  \tablenum[table-format=-1.4]{0.0027} &  \tablenum[table-format=-1.4]{-0.1135} \\
      H  & \tablenum[table-format=-1.4]{0.0000} &  \tablenum[table-format=-1.4]{-0.7614} &  \tablenum[table-format=-1.4]{0.4363} \\
      H  & \tablenum[table-format=-1.4]{0.0000} &  \tablenum[table-format=-1.4]{0.7398} &  \tablenum[table-format=-1.4]{0.4720}    
    \end{tabular}
\end{table}
\clearpage

\subsection{Optimized \ce{H2O} coupled to two orthogonal cavity modes}

All optimizations were performed using the \gls{bf} method, starting from the exact Hessian matrix in the first step and recalculating it after every 10th step.
The minima found were verified to have no imaginary frequency.

\begin{table}
  \caption{$\lambda_m = \SI{0.015}{\au}$, $q_1 = \SI{-1.5005}{\au}$ and $q_2 = \SI{0.0000}{\au}$.}
    \centering
    \begin{tabular}{ c ccc}       
    \hline
      O  & \tablenum[table-format=-1.4]{-0.1136} &  \tablenum[table-format=-1.4]{0.0000} &  \tablenum[table-format=-1.4]{0.0000} \\
      H  & \tablenum[table-format=-1.4]{0.4542} &  \tablenum[table-format=-1.4]{0.0000} &  \tablenum[table-format=-1.4]{0.7535} \\
      H  & \tablenum[table-format=-1.4]{0.4542} &  \tablenum[table-format=-1.4]{0.0000} &  \tablenum[table-format=-1.4]{-0.7535}    
    \end{tabular}
\end{table}

\begin{table}
  \caption{$\lambda_m = \SI{0.031}{\au}$, $q_1 = \SI{-3.0012}{\au}$ and $q_2 = \SI{0.0000}{\au}$.}
    \centering
    \begin{tabular}{ c ccc}       
    \hline
      O  & \tablenum[table-format=-1.4]{-0.1135} &  \tablenum[table-format=-1.4]{0.0000} &  \tablenum[table-format=-1.4]{0.0000} \\
      H  & \tablenum[table-format=-1.4]{0.4536} &  \tablenum[table-format=-1.4]{0.0000} &  \tablenum[table-format=-1.4]{0.7535} \\
      H  & \tablenum[table-format=-1.4]{0.4536} &  \tablenum[table-format=-1.4]{0.0000} &  \tablenum[table-format=-1.4]{-0.7535}    
    \end{tabular}
\end{table}

\begin{table}
  \caption{$\lambda_m = \SI{0.046}{\au}$, $q_1 = \SI{-4.5021}{\au}$ and $q_2 = \SI{0.0000}{\au}$.}
    \centering
    \begin{tabular}{ c ccc}       
    \hline
      O  & \tablenum[table-format=-1.4]{-0.1132} &  \tablenum[table-format=-1.4]{0.0000} &  \tablenum[table-format=-1.4]{0.0000} \\
      H  & \tablenum[table-format=-1.4]{0.4526} &  \tablenum[table-format=-1.4]{0.0000} &  \tablenum[table-format=-1.4]{0.7536} \\
      H  & \tablenum[table-format=-1.4]{0.4526} &  \tablenum[table-format=-1.4]{0.0000} &  \tablenum[table-format=-1.4]{-0.7536}    
    \end{tabular}
\end{table}

\begin{table}
  \caption{$\lambda_m = \SI{0.062}{\au}$, $q_1 = \SI{-6.0033}{\au}$ and $q_2 = \SI{0.0000}{\au}$.}
    \centering
    \begin{tabular}{ c ccc}       
    \hline
      O  & \tablenum[table-format=-1.4]{-0.1129} &  \tablenum[table-format=-1.4]{0.0000} &  \tablenum[table-format=-1.4]{0.0000} \\
      H  & \tablenum[table-format=-1.4]{0.4513} &  \tablenum[table-format=-1.4]{0.0000} &  \tablenum[table-format=-1.4]{0.7537} \\
      H  & \tablenum[table-format=-1.4]{0.4513} &  \tablenum[table-format=-1.4]{0.0000} &  \tablenum[table-format=-1.4]{-0.7537}    
    \end{tabular}
\end{table}

\begin{table}
  \caption{$\lambda_m = \SI{0.077}{\au}$, $q_1 = \SI{-7.5050}{\au}$ and $q_2 = \SI{0.0000}{\au}$.}
    \centering
    \begin{tabular}{ c ccc}       
    \hline
      O  & \tablenum[table-format=-1.4]{-0.1125} &  \tablenum[table-format=-1.4]{0.0000} &  \tablenum[table-format=-1.4]{0.0000} \\
      H  & \tablenum[table-format=-1.4]{0.4497} &  \tablenum[table-format=-1.4]{0.0000} &  \tablenum[table-format=-1.4]{0.7538} \\
      H  & \tablenum[table-format=-1.4]{0.4497} &  \tablenum[table-format=-1.4]{0.0000} &  \tablenum[table-format=-1.4]{-0.7538}    
    \end{tabular}
\end{table}

\begin{table}
  \caption{$\lambda_m = \SI{0.093}{\au}$, $q_1 = \SI{-9.0068}{\au}$ and $q_2 = \SI{0.0000}{\au}$.}
    \centering
    \begin{tabular}{ c ccc}       
    \hline
      O  & \tablenum[table-format=-1.4]{-0.1120} &  \tablenum[table-format=-1.4]{0.0000} &  \tablenum[table-format=-1.4]{0.0000} \\
      H  & \tablenum[table-format=-1.4]{0.4480} &  \tablenum[table-format=-1.4]{0.0000} &  \tablenum[table-format=-1.4]{0.7538} \\
      H  & \tablenum[table-format=-1.4]{0.4480} &  \tablenum[table-format=-1.4]{0.0000} &  \tablenum[table-format=-1.4]{-0.7538}    
    \end{tabular}
\end{table}

\begin{table}
  \caption{$\lambda_m = \SI{0.101}{\au}$, $q_1 = \SI{-10.5084}{\au}$ and $q_2 = \SI{0.0000}{\au}$.}
    \centering
    \begin{tabular}{ c ccc}       
    \hline
      O  & \tablenum[table-format=-1.4]{-0.1114} &  \tablenum[table-format=-1.4]{0.0000} &  \tablenum[table-format=-1.4]{0.0000} \\
      H  & \tablenum[table-format=-1.4]{0.4453} &  \tablenum[table-format=-1.4]{0.0000} &  \tablenum[table-format=-1.4]{0.7542} \\
      H  & \tablenum[table-format=-1.4]{0.4460} &  \tablenum[table-format=-1.4]{0.0000} &  \tablenum[table-format=-1.4]{-0.7542}    
    \end{tabular}
\end{table}

\begin{table}
  \caption{$\lambda_m = \SI{0.123}{\au}$, $q_1 = \SI{-12.0095}{\au}$ and $q_2 = \SI{0.0000}{\au}$.}
    \centering
    \begin{tabular}{ c ccc}       
    \hline
      O  & \tablenum[table-format=-1.4]{-0.1108} &  \tablenum[table-format=-1.4]{0.0000} &  \tablenum[table-format=-1.4]{0.0000} \\
      H  & \tablenum[table-format=-1.4]{0.4430} &  \tablenum[table-format=-1.4]{0.0000} &  \tablenum[table-format=-1.4]{0.7543} \\
      H  & \tablenum[table-format=-1.4]{0.4430} &  \tablenum[table-format=-1.4]{0.0000} &  \tablenum[table-format=-1.4]{-0.7543}    
    \end{tabular}
\end{table}
\clearpage

\subsection{Optimized \ce{H2O2} coupled to a single cavity mode}

All optimizations were performed using the \gls{bf} method, starting from the exact Hessian matrix in the first step.
The minima found were verified to have no imaginary frequency.

\begin{table}
  \caption{$\lambda_m = \SI{0.017}{\au}$ and $q_1 = \SI{1.8231}{\au}$.}
    \centering
    \begin{tabular}{ c ccc}       
    \hline
      O  & \tablenum[table-format=-1.4]{-0.1693} &  \tablenum[table-format=-1.4]{-0.6737} &  \tablenum[table-format=-1.4]{0.0575} \\
      H  & \tablenum[table-format=-1.4]{0.5197} &  \tablenum[table-format=-1.4]{-1.0623} &  \tablenum[table-format=-1.4]{-0.4600} \\
      O  & \tablenum[table-format=-1.4]{0.1693} &  \tablenum[table-format=-1.4]{0.6737} &  \tablenum[table-format=-1.4]{0.0575} \\  
      H  & \tablenum[table-format=-1.4]{-0.5197} &  \tablenum[table-format=-1.4]{1.0623} &  \tablenum[table-format=-1.4]{-0.4600} 
    \end{tabular}
\end{table}

\begin{table}
  \caption{$\lambda_m = \SI{0.033}{\au}$ and $q_1 = \SI{3.5933}{\au}$.}
    \centering
    \begin{tabular}{ c ccc}       
    \hline
      O  & \tablenum[table-format=-1.4]{-0.1066} &  \tablenum[table-format=-1.4]{-0.6864} &  \tablenum[table-format=-1.4]{0.0566} \\
      H  & \tablenum[table-format=-1.4]{0.6204} &  \tablenum[table-format=-1.4]{-1.0101} &  \tablenum[table-format=-1.4]{-0.4529} \\
      O  & \tablenum[table-format=-1.4]{0.1066} &  \tablenum[table-format=-1.4]{0.6864} &  \tablenum[table-format=-1.4]{0.0566} \\  
      H  & \tablenum[table-format=-1.4]{-0.6204} &  \tablenum[table-format=-1.4]{1.0101} &  \tablenum[table-format=-1.4]{-0.4529} 
    \end{tabular}
\end{table}

\begin{table}
  \caption{$\lambda_m = \SI{0.050}{\au}$ and $q_1 = \SI{5.2569}{\au}$.}
    \centering
    \begin{tabular}{ c ccc}       
    \hline
      O  & \tablenum[table-format=-1.4]{-0.1731} &  \tablenum[table-format=-1.4]{-0.6727} &  \tablenum[table-format=-1.4]{0.0551} \\
      H  & \tablenum[table-format=-1.4]{0.5276} &  \tablenum[table-format=-1.4]{-1.0669} &  \tablenum[table-format=-1.4]{-0.4406} \\
      O  & \tablenum[table-format=-1.4]{0.1731} &  \tablenum[table-format=-1.4]{0.6727} &  \tablenum[table-format=-1.4]{0.0551} \\  
      H  & \tablenum[table-format=-1.4]{-0.5276} &  \tablenum[table-format=-1.4]{1.0669} &  \tablenum[table-format=-1.4]{-0.4406} 
    \end{tabular}
\end{table}

\begin{table}
  \caption{$\lambda_m = \SI{0.067}{\au}$ and $q_1 = \SI{6.7444}{\au}$.}
    \centering
    \begin{tabular}{ c ccc}       
    \hline
      O  & \tablenum[table-format=-1.4]{-0.1766} &  \tablenum[table-format=-1.4]{-0.6719} &  \tablenum[table-format=-1.4]{0.0529} \\
      H  & \tablenum[table-format=-1.4]{0.5342} &  \tablenum[table-format=-1.4]{-1.0711} &  \tablenum[table-format=-1.4]{-0.4229} \\
      O  & \tablenum[table-format=-1.4]{0.1766} &  \tablenum[table-format=-1.4]{0.6719} &  \tablenum[table-format=-1.4]{0.0529} \\  
      H  & \tablenum[table-format=-1.4]{-0.5342} &  \tablenum[table-format=-1.4]{1.0711} &  \tablenum[table-format=-1.4]{-0.4229} 
    \end{tabular}
\end{table}

\begin{table}
  \caption{$\lambda_m = \SI{0.083}{\au}$ and $q_1 = \SI{8.7840}{\au}$.}
    \centering
    \begin{tabular}{ c ccc}       
    \hline
      O  & \tablenum[table-format=-1.4]{-0.1869} &  \tablenum[table-format=-1.4]{-0.6693} &  \tablenum[table-format=-1.4]{0.0456} \\
      H  & \tablenum[table-format=-1.4]{0.5535} &  \tablenum[table-format=-1.4]{-1.0840} &  \tablenum[table-format=-1.4]{-0.3644} \\
      O  & \tablenum[table-format=-1.4]{0.1869} &  \tablenum[table-format=-1.4]{0.6693} &  \tablenum[table-format=-1.4]{0.0456} \\  
      H  & \tablenum[table-format=-1.4]{-0.5535} &  \tablenum[table-format=-1.4]{1.0840} &  \tablenum[table-format=-1.4]{-0.3644} 
    \end{tabular}
\end{table}

\begin{table}
  \caption{$\lambda_m = \SI{0.100}{\au}$ and $q_1 = \SI{8.9450}{\au}$.}
    \centering
    \begin{tabular}{ c ccc}       
    \hline
      O  & \tablenum[table-format=-1.4]{-0.1944} &  \tablenum[table-format=-1.4]{-0.6675} &  \tablenum[table-format=-1.4]{0.0396} \\
      H  & \tablenum[table-format=-1.4]{0.5662} &  \tablenum[table-format=-1.4]{-1.0934} &  \tablenum[table-format=-1.4]{-0.3169} \\
      O  & \tablenum[table-format=-1.4]{0.1944} &  \tablenum[table-format=-1.4]{0.6675} &  \tablenum[table-format=-1.4]{0.0396} \\  
      H  & \tablenum[table-format=-1.4]{-0.5662} &  \tablenum[table-format=-1.4]{1.0934} &  \tablenum[table-format=-1.4]{-0.3169} 
    \end{tabular}
\end{table}

\begin{table}
  \caption{$\lambda_m = \SI{0.117}{\au}$ and $q_1 = \SI{7.7993}{\au}$.}
    \centering
    \begin{tabular}{ c ccc}       
    \hline
      O  & \tablenum[table-format=-1.4]{-0.2030} &  \tablenum[table-format=-1.4]{-0.6656} &  \tablenum[table-format=-1.4]{0.0301} \\
      H  & \tablenum[table-format=-1.4]{0.5841} &  \tablenum[table-format=-1.4]{-1.1051} &  \tablenum[table-format=-1.4]{-0.2407} \\
      O  & \tablenum[table-format=-1.4]{0.2030} &  \tablenum[table-format=-1.4]{0.6656} &  \tablenum[table-format=-1.4]{0.0301} \\  
      H  & \tablenum[table-format=-1.4]{-0.5841} &  \tablenum[table-format=-1.4]{1.1051} &  \tablenum[table-format=-1.4]{-0.2407} 
    \end{tabular}
\end{table}

\begin{table}
  \caption{$\lambda_m = \SI{0.133}{\au}$ and $q_1 = \SI{0.0007}{\au}$.}
    \centering
    \begin{tabular}{ c ccc}       
    \hline
      O  & \tablenum[table-format=-1.4]{-0.2124} &  \tablenum[table-format=-1.4]{-0.6638} &  \tablenum[table-format=-1.4]{0.0251} \\
      H  & \tablenum[table-format=-1.4]{0.6112} &  \tablenum[table-format=-1.4]{-1.1193} &  \tablenum[table-format=-1.4]{-0.1507} \\
      O  & \tablenum[table-format=-1.4]{0.2124} &  \tablenum[table-format=-1.4]{0.6638} &  \tablenum[table-format=-1.4]{0.0251} \\  
      H  & \tablenum[table-format=-1.4]{-0.6112} &  \tablenum[table-format=-1.4]{1.1193} &  \tablenum[table-format=-1.4]{-0.1507} 
    \end{tabular}
\end{table}

\begin{table}
  \caption{$\lambda_m = \SI{0.150}{\au}$ and $q_1 = \SI{0.0000}{\au}$.}
    \centering
    \begin{tabular}{ c ccc}       
    \hline
      O  & \tablenum[table-format=-1.4]{-0.2130} &  \tablenum[table-format=-1.4]{-0.6631} &  \tablenum[table-format=-1.4]{0.0000} \\
      H  & \tablenum[table-format=-1.4]{0.6097} &  \tablenum[table-format=-1.4]{-1.1189} &  \tablenum[table-format=-1.4]{0.0000} \\
      O  & \tablenum[table-format=-1.4]{0.2130} &  \tablenum[table-format=-1.4]{0.6097} &  \tablenum[table-format=-1.4]{0.0000} \\  
      H  & \tablenum[table-format=-1.4]{-0.5197} &  \tablenum[table-format=-1.4]{1.1189} &  \tablenum[table-format=-1.4]{0.0000} 
    \end{tabular}
\end{table}

\clearpage

\subsection{Optimized \ce{H2O2} coupled to two orthogonal cavity modes}

All optimizations were performed using the \gls{bf} method, starting from the exact Hessian matrix in the first step.
The minima found were verified to have no imaginary frequency.

\begin{table}
  \caption{$\lambda_m = \SI{0.017}{\au}$, $q_1 = \SI{1.8296}{\au}$ and $q_2 = \SI{0.0000}{\au}$.}
    \centering
    \begin{tabular}{ c ccc}       
    \hline
      O  & \tablenum[table-format=-1.4]{-0.2056} &  \tablenum[table-format=-1.4]{-0.6635} &  \tablenum[table-format=-1.4]{0.0577} \\
      H  & \tablenum[table-format=-1.4]{0.4597} &  \tablenum[table-format=-1.4]{-1.0887} &  \tablenum[table-format=-1.4]{-0.4618} \\
      O  & \tablenum[table-format=-1.4]{0.2056} &  \tablenum[table-format=-1.4]{0.6635} &  \tablenum[table-format=-1.4]{0.0577} \\  
      H  & \tablenum[table-format=-1.4]{-0.4597} &  \tablenum[table-format=-1.4]{1.0887} &  \tablenum[table-format=-1.4]{-0.4618} 
    \end{tabular}
\end{table}

\begin{table}
  \caption{$\lambda_m = \SI{0.033}{\au}$, $q_1 = \SI{3.6526}{\au}$ and $q_2 = \SI{0.0000}{\au}$.}
    \centering
    \begin{tabular}{ c ccc}       
    \hline
      O  & \tablenum[table-format=-1.4]{-0.2060} &  \tablenum[table-format=-1.4]{-0.6630} &  \tablenum[table-format=-1.4]{0.0575} \\
      H  & \tablenum[table-format=-1.4]{0.4591} &  \tablenum[table-format=-1.4]{-1.0894} &  \tablenum[table-format=-1.4]{-0.4602} \\
      O  & \tablenum[table-format=-1.4]{0.2060} &  \tablenum[table-format=-1.4]{0.6630} &  \tablenum[table-format=-1.4]{0.0575} \\  
      H  & \tablenum[table-format=-1.4]{-0.4591} &  \tablenum[table-format=-1.4]{1.0894} &  \tablenum[table-format=-1.4]{-0.4602} 
    \end{tabular}
\end{table}

\begin{table}
  \caption{$\lambda_m = \SI{0.050}{\au}$, $q_1 = \SI{5.4627}{\au}$ and $q_2 = \SI{0.0000}{\au}$.}
    \centering
    \begin{tabular}{ c ccc}       
    \hline
      O  & \tablenum[table-format=-1.4]{-0.2067} &  \tablenum[table-format=-1.4]{-0.6623} &  \tablenum[table-format=-1.4]{0.0572} \\
      H  & \tablenum[table-format=-1.4]{0.4581} &  \tablenum[table-format=-1.4]{-1.0904} &  \tablenum[table-format=-1.4]{-0.4575} \\
      O  & \tablenum[table-format=-1.4]{0.2067} &  \tablenum[table-format=-1.4]{0.6623} &  \tablenum[table-format=-1.4]{0.0572} \\  
      H  & \tablenum[table-format=-1.4]{-0.4581} &  \tablenum[table-format=-1.4]{1.0904} &  \tablenum[table-format=-1.4]{-0.4575} 
    \end{tabular}
\end{table}

\begin{table}
  \caption{$\lambda_m = \SI{0.067}{\au}$, $q_1 = \SI{7.2555}{\au}$ and $q_2 = \SI{0.0000}{\au}$.}
    \centering
    \begin{tabular}{ c ccc}       
    \hline
      O  & \tablenum[table-format=-1.4]{-0.2076} &  \tablenum[table-format=-1.4]{-0.6612} &  \tablenum[table-format=-1.4]{0.0567} \\
      H  & \tablenum[table-format=-1.4]{0.4569} &  \tablenum[table-format=-1.4]{-1.0920} &  \tablenum[table-format=-1.4]{-0.4539} \\
      O  & \tablenum[table-format=-1.4]{0.2076} &  \tablenum[table-format=-1.4]{0.6612} &  \tablenum[table-format=-1.4]{0.0567} \\  
      H  & \tablenum[table-format=-1.4]{-0.4569} &  \tablenum[table-format=-1.4]{1.0920} &  \tablenum[table-format=-1.4]{-0.4539} 
    \end{tabular}
\end{table}

\begin{table}
  \caption{$\lambda_m = \SI{0.083}{\au}$, $q_1 = \SI{9.0194}{\au}$ and $q_2 = \SI{0.0000}{\au}$.}
    \centering
    \begin{tabular}{ c ccc}       
    \hline
      O  & \tablenum[table-format=-1.4]{-0.2085} &  \tablenum[table-format=-1.4]{-0.6600} &  \tablenum[table-format=-1.4]{0.0562} \\
      H  & \tablenum[table-format=-1.4]{0.4555} &  \tablenum[table-format=-1.4]{-1.0938} &  \tablenum[table-format=-1.4]{-0.4493} \\
      O  & \tablenum[table-format=-1.4]{0.2085} &  \tablenum[table-format=-1.4]{0.6600} &  \tablenum[table-format=-1.4]{0.0562} \\  
      H  & \tablenum[table-format=-1.4]{-0.4555} &  \tablenum[table-format=-1.4]{1.0938} &  \tablenum[table-format=-1.4]{-0.4493} 
    \end{tabular}
\end{table}

\begin{table}
  \caption{$\lambda_m = \SI{0.100}{\au}$, $q_1 = \SI{10.7484}{\au}$ and $q_2 = \SI{0.0000}{\au}$.}
    \centering
    \begin{tabular}{ c ccc}       
    \hline
      O  & \tablenum[table-format=-1.4]{-0.2097} &  \tablenum[table-format=-1.4]{-0.6585} &  \tablenum[table-format=-1.4]{0.0555} \\
      H  & \tablenum[table-format=-1.4]{0.4538} &  \tablenum[table-format=-1.4]{-1.0961} &  \tablenum[table-format=-1.4]{-0.4437} \\
      O  & \tablenum[table-format=-1.4]{0.2097} &  \tablenum[table-format=-1.4]{0.6585} &  \tablenum[table-format=-1.4]{0.0555} \\  
      H  & \tablenum[table-format=-1.4]{-0.4538} &  \tablenum[table-format=-1.4]{1.0961} &  \tablenum[table-format=-1.4]{-0.4437} 
    \end{tabular}
\end{table}

\begin{table}
  \caption{$\lambda_m = \SI{0.117}{\au}$, $q_1 = \SI{12.4338}{\au}$ and $q_2 = \SI{0.0000}{\au}$.}
    \centering
    \begin{tabular}{ c ccc}       
    \hline
      O  & \tablenum[table-format=-1.4]{-0.2111} &  \tablenum[table-format=-1.4]{-0.6568} &  \tablenum[table-format=-1.4]{0.0547} \\
      H  & \tablenum[table-format=-1.4]{0.4519} &  \tablenum[table-format=-1.4]{-1.0988} &  \tablenum[table-format=-1.4]{-0.4372} \\
      O  & \tablenum[table-format=-1.4]{0.2111} &  \tablenum[table-format=-1.4]{0.6568} &  \tablenum[table-format=-1.4]{0.0547} \\  
      H  & \tablenum[table-format=-1.4]{-0.4519} &  \tablenum[table-format=-1.4]{1.0988} &  \tablenum[table-format=-1.4]{-0.4372} 
    \end{tabular}
\end{table}

\begin{table}
  \caption{$\lambda_m = \SI{0.133}{\au}$, $q_1 = \SI{14.0671}{\au}$ and $q_2 = \SI{0.0000}{\au}$.}
    \centering
    \begin{tabular}{ c ccc}       
    \hline
      O  & \tablenum[table-format=-1.4]{-0.2126} &  \tablenum[table-format=-1.4]{-0.6548} &  \tablenum[table-format=-1.4]{0.0537} \\
      H  & \tablenum[table-format=-1.4]{0.4495} &  \tablenum[table-format=-1.4]{-1.1018} &  \tablenum[table-format=-1.4]{-0.4300} \\
      O  & \tablenum[table-format=-1.4]{0.2126} &  \tablenum[table-format=-1.4]{0.6548} &  \tablenum[table-format=-1.4]{0.0537} \\  
      H  & \tablenum[table-format=-1.4]{-0.4495} &  \tablenum[table-format=-1.4]{1.1018} &  \tablenum[table-format=-1.4]{-0.4300} 
    \end{tabular}
\end{table}

\begin{table}
  \caption{$\lambda_m = \SI{0.150}{\au}$, $q_1 = \SI{15.6389}{\au}$ and $q_2 = \SI{0.0000}{\au}$.}
    \centering
    \begin{tabular}{ c ccc}       
    \hline
      O  & \tablenum[table-format=-1.4]{-0.2144} &  \tablenum[table-format=-1.4]{-0.6526} &  \tablenum[table-format=-1.4]{0.0527} \\
      H  & \tablenum[table-format=-1.4]{0.4468} &  \tablenum[table-format=-1.4]{-1.1053} &  \tablenum[table-format=-1.4]{-0.4219} \\
      O  & \tablenum[table-format=-1.4]{0.2144} &  \tablenum[table-format=-1.4]{0.6526} &  \tablenum[table-format=-1.4]{0.0527} \\  
      H  & \tablenum[table-format=-1.4]{-0.4468} &  \tablenum[table-format=-1.4]{1.1053} &  \tablenum[table-format=-1.4]{-0.4219}
    \end{tabular}
\end{table}

\clearpage

\bibliography{lit.bib}
\providecommand{\latin}[1]{#1}
\makeatletter
\providecommand{\doi}
  {\begingroup\let\do\@makeother\dospecials
  \catcode`\{=1 \catcode`\}=2 \doi@aux}
\providecommand{\doi@aux}[1]{\endgroup\texttt{#1}}
\makeatother
\providecommand*\mcitethebibliography{\thebibliography}
\csname @ifundefined\endcsname{endmcitethebibliography}
  {\let\endmcitethebibliography\endthebibliography}{}